\makeatletter
\def\cl@chapter{}
\makeatother
\RequirePackage{fix-cm}
\documentclass[smallextended,natbib]{svjour3}       
\smartqed  
\RequirePackage{microtype}
\SetExtraKerning{
  encoding = {OT1,T1,T2A,LY1,OT4,QX,T5,TS1,EU1,EU2}
}{%
  \textemdash = {167,167},
  — = {167,167}
}
\usepackage[T1]{fontenc}
\usepackage[utf8]{inputenc}
\usepackage{amssymb}
\usepackage{amsmath}
\allowdisplaybreaks
\usepackage{caption}
\usepackage{subcaption}
\usepackage{url}
\usepackage{adjustbox}
\usepackage{lscape}

\usepackage{defs}
\graphicspath{{./img/}}

\newcommand{\numIterations}{30}
\newcommand{\numUniqueClasses}{134}
\newcommand{\numUniqueClassesWithAssertions}{105}
\newcommand{\MIOTypeHintsEffect}{0.5020264659290612}

\newcommand{\MOSATypeHintsEffect}{0.5021431400212866}

\newcommand{\RandomTypeHintsEffect}{0.5407002920719785}

\newcommand{\RandomTypeHintsPValue}{1.0820629628729027e-10}

\newcommand{\WSTypeHintsEffect}{0.5084451808123561}

\newcommand{\CoverageLocCorrelationDynaMOSAWithTypesPearsonR}{-0.21110153477989313}
\newcommand{\CoverageLocCorrelationDynaMOSAWithTypesPValue}{0.014347567754609247}
\newcommand{\CoverageMcCabeCorrelationDynaMOSAWithTypesPearsonR}{-0.36546487138562506}
\newcommand{\CoverageMcCabeCorrelationDynaMOSAWithTypesPValue}{2.996282119913574e-05}
\newcommand{\numProjects}{20}
\newcommand{\CoverageMutationScoreCorrelationDynaMOSAWithTypesPearsonR}{0.45010420483233965}
\newcommand{\CoverageMutationScoreCorrelationDynaMOSAWithTypesPValue}{5.530206831379375e-157}
\newcommand{\numTotalClasses}{163}

\newcommand{\CoverageImprovementsMIOBetter}{13}
\newcommand{\CoverageImprovementsMIOWorse}{3}
\newcommand{\CoverageImprovementsMOSABetter}{5}
\newcommand{\CoverageImprovementsMOSAWorse}{1}
\newcommand{\CoverageImprovementsRandomBetter}{52}
\newcommand{\CoverageImprovementsRandomWorse}{0}
\newcommand{\CoverageImprovementsWSBetter}{25}
\newcommand{\CoverageImprovementsWSWorse}{1}
\newcommand{\MedianCovDynaMOSATypeHints}{80.7207498383969}

\journalname{Empirical Software Engineering}
\begin{document}

\title{An Empirical Study of Automated Unit Test Generation for Python}

\author{Stephan Lukasczyk \and
  Florian Kroiß \and
  Gordon Fraser}

\institute{S. Lukasczyk \at
              University of Passau \\
              Innstr. 33, 94032 Passau, Germany \\
              \email{stephan.lukasczyk@uni-passau.de}
           \and
           F. Kroiß \at
              University of Passau \\
              Innstr. 33, 94032 Passau, Germany \\
              \email{kroiss@fim.uni-passau.de}
          \and
          G. Fraser \at
              University of Passau \\
              Innstr. 33, 94032 Passau, Germany \\
              \email{gordon.fraser@uni-passau.de}
}

\date{Received: date / Accepted: date}

\maketitle

\begin{abstract}
  %
  Various mature automated test generation tools exist
  for statically typed programming languages such as Java.
  %
  Automatically generating unit tests for dynamically typed programming languages
  such as Python, however,
  is substantially more difficult due to the dynamic nature of these languages
  as well as the lack of type information.
  %
  Our \pynguin framework provides automated unit test generation for Python.
  %
  %
  %
  In this paper,
  we extend our previous work on \pynguin
  to support more aspects of the Python language,
  and by studying a larger variety of well-established
  state of the art test-generation algorithms, namely DynaMOSA, MIO, and MOSA.
  Furthermore,
  we improved our \pynguin tool to generate regression assertions,
  whose quality we also evaluate.
  %
  Our experiments confirm that evolutionary algorithms
  can outperform random test generation also in the context of Python, and
  similar to the Java world,
  DynaMOSA yields the highest coverage results.
  However,
  our results also demonstrate that there are still fundamental remaining issues,
  such as inferring type information for code without this information,
  currently limiting the effectiveness of test generation for Python.

\end{abstract}

\section{Introduction}\label{sec:introduction}

%
Automated unit test generation is an established
field in research
and a technique well-received by researchers and practitioners
to support programmers.
Mature research prototypes exist,
implementing test-generation approaches
such as feedback-directed random generation~\citep{PLE+07}
or evolutionary algorithms~\citep{CGA+18}.
These techniques enable the automated generation of unit tests
for statically typed programming languages,
such as Java,
and remove the burden of the potentially tedious task
of writing unit tests from the programmer.
%

%
In recent years,
however,
dynamically typed programming languages,
most notably JavaScript, Python, and Ruby,
have gained huge popularity amongst practitioners.
Current rankings,
such as the IEEE Spectrum Ranking\footnote{%
  \url{https://spectrum.ieee.org/at-work/tech-careers/top-programming-language-2020},
  accessed 2021–02–02.%
} underline this popularity,
now listing Python as the overall most popular language
according to their ranking criteria.
Dynamically typed languages
lack static type information on purpose as this is
%
supposed to enable rapid development~\citep{GPS+15}. However, this dynamic nature has also been reported to cause
reduced development productivity~\citep{KHR+12},
code usability~\citep{MHR+12},
or code quality~\citep{MR13,GBB17}.
The lack of a (visible static) type system is the main reason
for type errors encountered in these dynamic languages~\citep{XLZ+16}.
%

%
The lack of static types is particularly problematic for automated test
generation, which requires type information to provide appropriate parameter
types for method calls or to assemble complex objects.
In absence of type information
the test generation can only guess,
for example,
the appropriate parameter types for new function calls.
To overcome this limitation,
existing test generators for dynamically typed languages
often do not target test-generation for general APIs,
but resort to other means
such as using the document object model of a web browser
to generate tests for JavaScript~\citep{MMP15},
or by targeting specific systems,
such as the browser's event handling system~\citep{ADJ+2011,LAG14}.
A general purpose unit test generator at the API level
has not been available until recently.
%

%
\pynguin~\citep{LKF20,LF22} aims to fill this fundamental gap:
\pynguin is an automated unit test generation framework
for Python programs.
It takes a Python module as input together with the module's dependencies,
and aims to automatically generate unit tests
that maximise code coverage.
The version of \pynguin we evaluated in our previous work~\citep{LKF20}
implemented two established test generation techniques:
whole-suite generation~\citep{FA13}
and feedback-directed random generation~\citep{PLE+07}.
Our empirical evaluation showed that the whole-suite approach
is able to achieve higher code coverage than random testing.
Furthermore,
we studied the influence of available type information,
which leads to higher resulting coverage
if it can be utilised by the test generator.
%

%
In this paper we extend our previous evaluation~\citep{LKF20} in several aspects
and make the following contributions:
We implemented further test generation algorithms:
the many-objective sorting algorithm~(MOSA)~\citep{PKT15}
and its extension DynaMOSA~\citep{PKT18},
as well as the many independent objective~(MIO) algorithm~\citep{Arc17,Arc18}.
We study and compare the performance of all algorithms
in terms of resulting branch coverage.
We furthermore enlarge the corpus of subjects for test generation
to get a broader insight into test generation for Python.
For this we take projects from the \toolname{BugsInPy} project~\citep{WSL+20}
and the \toolname{ManyTypes4Py} dataset~\citep{MLG21};
in total, we add eleven new projects to our dataset.
Additionally,
we implemented the generation of regression assertions
based on mutation testing~\citep{FZ12}.
Our empirical evaluation
confirms that, similar to prior findings on Java,
DynaMOSA and MOSA perform best on Python code,
with a median of \SI{\MedianCovDynaMOSATypeHints}{\percent} branch coverage.
Furthermore,
our results show that type information
is an important contributor to coverage
on all studied test-generation algorithms,
although this effect is only significant for some of our subject systems,
showing an average impact of \SIrange{2.2}{4.3}{\percent} on coverage.
While this confirms that type information is one of the challenges
in testing Python code,
these results also suggest there are other open challenges for future research,
such as dealing with type hierarchies and constructing complex objects.

\section{Background}\label{sec:background}

The main approaches to automatically generate unit tests
are either by creating random sequences,
or by applying meta-heuristic search algorithms.
Random testing assembles sequences of calls to constructors and methods randomly,
often with the objective to find undeclared exceptions~\citep{CS04}
or violations of general object contracts~\citep{PLE+07},
but the generated tests can also be used as automated regression tests.
The effectiveness of random test generators can be increased
by integrating heuristics~\citep{MAZ+15,SPG15}.
Search-based approaches use a similar representation,
but apply evolutionary search algorithms
to maximize code coverage~\citep{Ton04,BM10,FA13,AML11}.

As an example to illustrate how type information is used by existing test
generators, consider the following snippets
of Java (left) and Python (right) code:
\begin{minipage}[t]{0.48\textwidth}
  \begin{lstlisting}[language=Java]
class Foo {
  Foo(Bar b) { ... }
  void doFoo(Bar b) { ... } }
class Bar {
  Bar() { ... }
  Bar doBar(Bar b) { ... } }
  \end{lstlisting}
\end{minipage}%
\hfill%
\begin{minipage}[t]{0.48\textwidth}
  \begin{lstlisting}[language=Python]
class Foo:
  def __init__(self, b): ...
  def do_foo(self, b): ...
class Bar:
  def __init__(self): ...
  def do_bar(self, b): ...
  \end{lstlisting}
\end{minipage}

Assume \mintinline{java}{Foo}
of the Java example
is the class under test. It has a dependency on
class \mintinline{java}{Bar}: in order to generate an object of type
\mintinline{java}{Foo} we need an instance of \mintinline{java}{Bar}, and the
method \mintinline{java}{doFoo} also requires a parameter of type
\mintinline{java}{Bar}.

Random test generation would typically generate tests in a forward way.
Starting with an empty sequence $t_0 = \langle \rangle$, all available calls
for which all parameters can be satisfied with objects already existing in the
sequence can be selected. In our example, initially only the constructor of
\mintinline{java}{Bar} can be called, since all other methods and constructors
require a parameter, resulting in $t_1 = \langle o_1 = $ \mintinline{java}{new
Bar()}$\rangle$. Since
$t_1$ contains an object of type \mintinline{java}{Bar}, in the second step the
test generator now has a choice of either invoking \mintinline{java}{doBar} on
that object (and use the same object also as parameter), or invoking the
constructor of \mintinline{java}{Foo}. Assuming the chosen call is the
constructor of \mintinline{java}{Foo}, we now have $t_2 = \langle o_1 = $
\mintinline{java}{new Bar()}$; o_2 = $ \mintinline{java}{new
Foo(}$o_1$\mintinline{java}{)}$; \rangle$. Since there now is also an instance
of
\mintinline{java}{Foo} in the sequence, in the next step also the method
\mintinline{java}{doFoo} is an option. The random test generator will continue
extending the sequence in this manner, possibly integrating heuristics to
select more relevant calls, or to decide when to start with a new sequence.

An alternative approach, for example applied during the mutation step of an
evolutionary test generator, is to select necessary calls in a backwards
fashion. That is, a search-based test generator like \evosuite~\citep{FA13}
would first decide that it needs to, for example, call method
\mintinline{java}{doFoo} of class \mintinline{java}{Foo}. In order to achieve
this, it requires an instance of \mintinline{java}{Foo} and an instance of
\mintinline{java}{Bar} to satisfy the dependencies. To generate a parameter
object of type \mintinline{java}{Bar}, the test generator would consider all
calls that are declared to return an instance of
\mintinline{java}{Bar}---which is the case for the constructor of
\mintinline{java}{Bar} in our example, so it would prepend a call to \mintinline{java}{Bar()}
before the invocation of \mintinline{java}{doFoo}. Furthermore, it would try to
instantiate \mintinline{java}{Foo} by calling the constructor. This, in turn,
requires an instance of \mintinline{java}{Bar}, for which the test generator
might use the existing instance, or could invoke the constructor of
\mintinline{java}{Bar}.

In both scenarios, type information is crucial: In the forward construction
type information is used to inform the choice of call to append to the
sequence, while in the backward construction type information is used to select
generators of dependency objects.
Without type information,
which is the case with the Python example,
a forward construction (1) has to allow all possible functions at all steps,
thus may not only select the constructor of \mintinline{python}{Bar},
but also that of \mintinline{python}{Foo} with an arbitrary parameter type,
and (2) has to consider all existing objects for all parameters of a selected
call, and thus, for example, also \mintinline{python}{str} or \mintinline{python}{int}.
Backwards construction without type information would also have to try to
select generators from all possible calls, and all possible objects,
which both result in a potentially large search space to select from.
Without type information in our example
we might see instantiations of \mintinline{python}{Foo}
or calls to \mintinline{python}{do\_foo}
with parameter values of unexpected types, such as:
\begin{center}
  \begin{lstlisting}[language=python]
var_1 = Bar()
var_2 = Foo(42)
var_3 = var_2.do_foo("hello")
  \end{lstlisting}
\end{center}
This example violates the assumptions of the programmer,
which are that the constructor of \mintinline{python}{Foo}
and the \mintinline{python}{do\_foo} method both expect an object
of type \mintinline{python}{Bar}.
When type information is not present
such test cases can be generated
and will only fail if the execution raises an exception;
for example,
due to a call to a method
that is defined for the \mintinline{python}{Bar} type
but not on an \mintinline{python}{int} or \mintinline{python}{str} object.
Type information can be provided in two ways in recent Python versions:
either in a stub file that contains type hints
or directly annotated in the source code.
A stub file can be compared to C header files:
they contain,
for example,
method declarations with their according types.
Since Python~3.5,
the types can also be annotated directly in the implementing source code,
in a similar fashion known from statically typed languages~(see
PEP~484\footnote{%
  \url{https://python.org/dev/peps/pep-0484/}, accessed 2021–05–04.%
  \label{foot:pep-484}
}).

\section{Unit Test Generation with \pynguin}\label{sec:approach}

We introduce our automated test-generation framework
for Python,
called \pynguin,
in the following sections.
We start with a general overview
on \pynguin.
Afterwards,
we formally introduce a representation for the test-generation problem
using evolutionary algorithms in Python.
We also discuss different components and operators
that we use.
%


\subsection{The \pynguin Framework}\label{sec:approach-algorithms}

\pynguin is a framework for automated unit test generation
written in and for the Python programming language.
The framework is available as open-source software
licensed under the GNU Lesser General Public License
from its GitHub repository\footnote{%
  \url{https://github.com/se2p/pynguin}, accessed 2022–07–06.%
}.
It can also be installed from the Python Package Index~(PyPI)\footnote{%
  \url{https://pypi.org/project/pynguin/}, accessed 2022–07–06.%
} using the \texttt{pip} utility.
We refer the reader to \pynguin's web site\footnote{%
  \url{https://www.pynguin.eu}, accessed 2022–07–06.%
} for further links and information on the tool.
%

\begin{figure}[th]
  \centering
  \includegraphics[width=\textwidth]{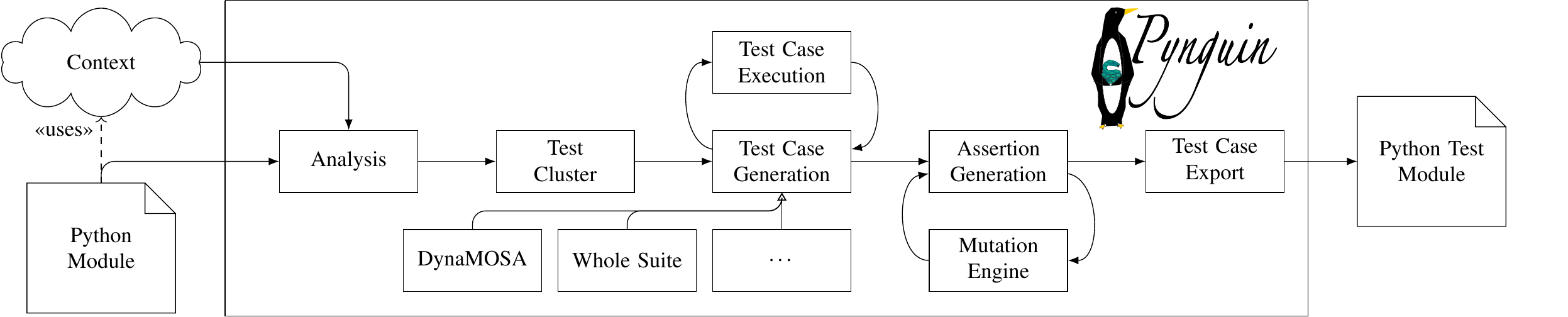}
  \caption{\label{fig:pynguin-components}%
    The different components of \pynguin~(taken from~\citet{LF22})%
  }
\end{figure}

\pynguin takes as input a Python module and allows the generation of unit tests
using different techniques.
For this,
it analyses the module
and extracts information about available methods from the module
and types from the module and its transitive dependencies
to build a test cluster~(see \cref{sec:approach-testcluster}).
Next,
it generates test cases using a variety of different
test-generation algorithms~(which we describe in the following sub-sections).
Afterwards,
it generates regression assertions
for the previously generated test cases~(see
\cref{sec:approach-assertion-generation}).
Finally,
a tool run emits the generated test cases
in the style of the widely-used \toolname{PyTest}\footnote{%
  \url{https://www.pytest.org}, accessed 2022–07–06.%
} framework.
\Cref{fig:pynguin-components} illustrates \pynguin's components
and their relationships.
For a detailed description of the components,
how to use \pynguin for own purposes,
as well as how to extend \pynguin,
we refer the reader to our respective tool paper~\citep{LF22}.

\pynguin is built to be extensible
with other test generation approaches and algorithms.
It comes with a variety of well-received test-generation algorithms,
which we describe in the following sections.

\subsubsection{Feedback-Directed Random Test Generation}\label{sec:approach-randoop}

\pynguin provides a feedback-directed random algorithm
adopted from \toolname{Randoop}~\citep{PLE+07}.
The algorithm starts with two empty test suites,
one for passing and one for failing tests,
and randomly adds statements to an empty test case.
The test case is executed after each addition.
Test cases that do not raise an exception are added to the passing test suite;
otherwise they are added to the failing test suite.
The algorithm will then randomly choose a test case
from the passing test suite
or an empty test case
as the basis to add further statements.
\Cref{alg:randoop} shows the basic algorithm.

\begin{algorithm}[t]
  \caption{\label{alg:randoop}Feedback-directed random generation~(adapted
  from~\citet{PLE+07})}

  \SetKwInput{Input}{Input}
  \SetKwInput{Output}{Output}
  \SetKw{Continue}{Continue}

  \Input{Stopping condition~\(C\), a set of modules~\(c\)}
  \Output{A pair of non-error sequences~\(s_n\) and error sequences~\(s_e\)}
  \( s_e \leftarrow \{\} \)\;
  \( s_n \leftarrow \{\} \)\;
  \While{\( \neg C \)}{%
    \( m(T_1\dots T_k) \leftarrow \textsc{RandomPublicMethod}(c) \)\;
    \( \langle S, V \rangle \leftarrow \textsc{RandomSeqsAndVals}(s_n, T_1\dots
    T_k \)\;
    \( N \leftarrow \textsc{Extend}(m, S, V) \)\;
    \If{\( N \in s_n \cup s_e \)}{%
      \Continue
    }
    \( \langle \vec{\mathbf{o}}, v \rangle \leftarrow \textsc{Execute}(N) \)\;
    \eIf{\( v \)}{%
      \( s_e \leftarrow s_e \cup \{N\} \)\;
    }{%
      \( s_n \leftarrow s_n \cup \{N\} \)\;
    }
  }
  \Return{\( \langle s_n, s_e \rangle \)}

\end{algorithm}

The main differences of our implementation
compared to \toolname{Randoop} are
that our version of the algorithm does not check for contract violations,
does not implement any of \toolname{Randoop}'s filtering criteria,
and it does not require the user to provide a list of relevant classes,
functions, and methods.
Please note that our random algorithm,
in contrast to the following evolutionary algorithms,
does not use a fitness function to guide its random search process.
It only checks the coverage of the generated test suite
to stop early once the module under test is fully covered.
We will refer to this algorithm as \emph{Random} in the following.

\subsubsection{Whole Suite Test Generation}\label{sec:approach-ws}

The whole suite approach implements a genetic algorithm
that takes a test suite,
that is a set of test cases,
as an individual~\citep{FA13}.
It uses a monotonic genetic algorithm,
as shown in \cref{alg:monotonic-ga} as its basis.
Mutation and crossover operators can modify the test suite as well as its
contents, that is, the test cases.
It uses the sum of all branch distances as a fitness function.
We will refer to this algorithm as \emph{WS} in the following.

\begin{algorithm}[t]
  \caption{\label{alg:monotonic-ga}Monotonic Genetic Algorithm~(adopted
  from~\cite{CGA+18})}

  \SetKwInput{Input}{Input}
  \SetKwInput{Output}{Output}

  \Input{Stopping condition~\(C\), Fitness function~\(\delta\), Population
    size~\(p_s\), Selection function~\(s_f\), Crossover function~\(c_f\),
    Crossover probability~\(c_p\), Mutation function~\(m_f\), Mutation
    probability~\(m_p\)}
  \Output{Population of optimised individuals~\(P\)}
  \(P \leftarrow \textsc{GenerateRandomPopulation}(p_s)\)\;
  \(\textsc{PerformFitnessEvaluation}(\delta, P)\)\;
  \While{\(\neg C\)}{%
    \(N_P \leftarrow \{\} \cup \textsc{Elitism}(P)\)\;
    \While{\(|N_P| < p_s\)}{%
      \(p_1, p_2 \leftarrow \textsc{Selection}(s_f, P)\)\;
      \(o_1, o_2 \leftarrow \textsc{Crossover}(c_f, c_p, p_1, p_2)\)\;
      \(\textsc{Mutation}(m_f, m_p, o_1)\)\;
      \(\textsc{Mutation}(m_f, m_p, o_2)\)\;
      \(\textsc{PerformFitnessEvaluation}(\delta, o_1)\)\;
      \(\textsc{PerformFitnessEvaluation}(\delta, o_2)\)\;
      \eIf{\(\textsc{Best}(o_1, o_2) \text{ is better than } \textsc{Best}(p_1,
      p_2)\)}{%
        \(N_P \leftarrow N_P \cup \{o_1, o_2\}\)\;
      }{%
        \(N_P \leftarrow N_P \cup \{p_1, p_2\}\)\;
      }
    }
    \(P \leftarrow N_P\)\;
  }
  \Return{\(P\)}

\end{algorithm}

\subsubsection{Many-Objective Sorting Algorithm~(MOSA)}\label{sec:approach-mosa}

The \emph{Many-Objective Sorting Algorithm}~(MOSA,~\citep{PKT15}) is
an evolutionary algorithm specifically designed to foster branch coverage
and overcome some of the limitations of whole-suite generation.
MOSA considers branch coverage as a many-objective optimisation problem,
whereas previous approaches,
such as whole suite test generation,
combined the objectives into one single value.
It therefore assigns each branch its individual objective function.
The basic MOSA algorithm is shown in \cref{alg:mosa}.

\begin{algorithm}[t]
  \caption{\label{alg:mosa}Many-Objective Sorting Algorithm~(MOSA, adopted
  from~\citet{CGA+18})}

  \SetKwInput{Input}{Input}
  \SetKwInput{Output}{Output}

  \Input{Stopping condition~\(C\), Fitness function~\(\delta\), Population
  size~\(p_s\), Crossover function~\(c_f\), Crossover probability~\(c_p\),
  Mutation probability~\(m_p\)}
  \Output{Archive of optimised individuals~\(A\)}
  \(p \leftarrow 0\)\;
  \(N_p \leftarrow \textsc{GenerateRandomPopulation}(p_s)\)\;
  \(\textsc{PerformFitnessEvaluation}(\delta, N_p)\)\;
  \(A \leftarrow \{\}\)\;
  \While{\(\neg C\)}{%
    \(N_o \leftarrow \textsc{GenerateOffspring}(c_f, c_p, m_p, N_p)\)\;
    \(R_t \leftarrow N_p \cup N_o\)\;
    \(r \leftarrow 0\)\;
    \(F_r \leftarrow \textsc{PreferenceSorting}(R_t)\)\;
    \(N_{p+1} \leftarrow \{\}\)\;
    \While{\(|N_{p+1}| + |F_r| \leq p_s\)}{%
      \(\textsc{CalculateCrowdingDistance}(F_r)\)\;
      \(N_{p+1} \leftarrow N_{p+1} \cup F_r\)\;
      \(r \leftarrow r + 1\)\;
    }
    \(\textsc{DistanceCrowdingSort}(F_r)\)\;
    \(N_{p+1} \leftarrow N_{p+1} \cup F_r \text{ with size } p_s - |N_{p+1}|\)\;
    \(\textsc{UpdateArchive}(A, N_{p+1})\)\;
    \(p \leftarrow p + 1\)\;
  }
  \Return{\(A\)}

\end{algorithm}

MOSA starts with an initial random population
and evolves this population to improve its ability to cover more branches.
Earlier many-objective genetic algorithms suffer
from the so-called dominance resistance problem~\citep{LBB14}.
This means that the proportion of non-dominated solutions increases
exponentially with the number of goals to optimise.
As a consequence the search process degrades to a random one.
MOSA specifically targets this
by introducing a preference criterion to choose the optimisation targets
to focus the search only on the still relevant targets,
that is, the yet uncovered branches.
Furthermore, MOSA keeps a second population, called \emph{archive},
to store already found solutions,
that is,
test cases that cover branches.
This archive is built in a way that it automatically prefers
shorter test cases over longer for the same coverage goals.
We refer the reader to the literature for details on MOSA~\citep{PKT15}.
In the following we will refer to this algorithm as \emph{MOSA}.

\subsubsection{Dynamic Target Selection MOSA~(DynaMOSA)}\label{sec:approach-dynamosa}

\emph{DynaMOSA}~\citep{PKT18} is an extension to the original MOSA algorithm.
The novel contribution of DynaMOSA was to dynamically select the targets
for the optimisation.
This selection is done on the control-dependency hierarchy of statements.
Let us consider the Python code snippet from \cref{lst:control-dependency}.

\begin{minipage}[b]{0.47\textwidth}
  \begin{lstlisting}[%
%    caption={A Python snippet showing control-dependent conditions},%
%    label={lst:control-dependency},%
    language=Python,%
    numbers=left,%
  ]
if foo < 42:
    if bar == 23:
        do_something()
  \end{lstlisting}
  \captionof{lstlisting}{\label{lst:control-dependency}%
    A Python snippet showing control-dependent conditions.%
  }
\end{minipage}
\hfill
\begin{minipage}[b]{0.47\textwidth}
  \centering
  \begin{tikzpicture}[node distance=0.5cm]
    \node [shape=circle, draw] (one) {1};
    \node [shape=circle, draw] (two) [below right=of one] {2};
    \node [shape=circle, draw] (three) [below right=of two] {3};
    \draw [-{Latex[]}] (one) -- (two);
    \draw [-{Latex[]}] (two) -- (three);
  \end{tikzpicture}
  \captionof{figure}{\label{fig:cdg}%
    The control-dependence graph for the snippet
    in \cref{lst:control-dependency}.
  }
\end{minipage}

The condition in line~2 is control dependent on the condition in the first line.
This means, it can only be reached and covered,
if the condition in the first line is satisfied.
Thus, searching for an assignment for the variable \texttt{bar}
to fulfil the condition—and thus covering line~3—is not necessary
unless the condition in line~1 is fulfilled.
\Cref{fig:cdg} depicts the control-dependence graph.
DynaMOSA uses the control-dependencies,
which are found in the control-dependence graph of the respective code,
to determine which goals to select
for further optimisation.
To compute the control dependencies within \pynguin,
we generate a control-flow graph from the module under test's byte code
using the \texttt{bytecode}\footnote{
  \url{https://www.pypi.org/project/bytecode}, accessed 2022–07–14.%
} library;
we use standard algorithms to compute post-dominator tree
and control-dependence graph
from the control-flow graph~(see,
for example,
the work of \citet{FOW87} for such algorithms).
We refer the reader to the literature for details on DynaMOSA~\citep{PKT18}.
In the following we will refer to this algorithm as \emph{DynaMOSA}.

\subsubsection{Many Independent Objectives~(MIO)}\label{sec:approach-mio}

The \emph{Many Independent Objectives} Algorithm~(MIO,~\citep{Arc17})
targets some limitations of both the whole suite and the MOSA approach.
To do this, it combines the simplicity of a \((1+1)\)EA with feedback-directed
target selection, dynamic exploration/exploitation, and a dynamic population.
MIO also maintains an archive of tests,
were it keeps a different population of tests for each testing target,
for example, for each branch to cover.
\Cref{alg:mio} shows its main algorithm.
MIO was designed with the aim of overcoming some intrinsic limitations
of the Whole Suite or MOSA algorithms
that arise especially in system-level test generation.
Such systems can contain hundreds of thousands of objectives to cover,
for which a fixed-size population will most likely be not suitable;
hence, MIO uses a dynamic population.
It furthermore turns out
that exploration is good at the beginning of the search
whereas a focused exploitation is beneficial for better results in the end;
MIO addresses this insight by a dynamic exploration/exploitation control.
Lastly,
again addressing the large number of objectives and limited resources,
MIO selects the objectives to focus its search on
by using a feedback-directed sampling technique.
The literature~\citep{Arc17,Arc18} provides more details on MIO
to the interested reader.
We will refer to this algorithm as \emph{MIO} in the following.

\begin{algorithm}[t]
  \caption{\label{alg:mio}Many Independent Objective~(MIO) Algorithm~(adapted
  from~\citet{CGA+18})}

  \SetKwInput{Input}{Input}
  \SetKwInput{Output}{Output}

  \Input{Stopping condition~\(C\), Fitness function~\(\delta\), Population
    size~\(N\), Number of mutations~\(M\), Mutation function~\(m_f\), Mutation probability~\(m_p\),
    Probability of random sampling~\(R\), Start of focus search~\(F\)}
  \Output{Archive of optimised individuals~\(A\)}
  \(Z \leftarrow \textsc{SetOfEmptyPopulations}()\)\;
  \(A \leftarrow \{\}\)\;
  \(p \leftarrow null \)\;
  \(m \leftarrow 1\)\;
  \While{\(\neg C\)}{%
    \If{\(p \neq null \land m < M\)}{%
       \(p \leftarrow \textsc{Mutation}(m_f, m_p, p)\)\;
       \(m \leftarrow m + 1\)\;
  	}
    \ElseIf{\(p = null \lor R > \textsc{RANDOM}(0, 1)\)}{%
      \(p \leftarrow \textsc{GenerateRandomIndividual}()\)\;
      \(m \leftarrow 1\)\;
    }
	\Else{%
      \(p \leftarrow \textsc{SampleIndividual}(Z)\)\;
      \(p \leftarrow \textsc{Mutation}(m_f, m_p, p)\)\;
      \(m \leftarrow 1\)\;
    }
    \ForAll{\(t \in \textsc{ReachedTargets}(p)\)}{%
      \eIf{\(\textsc{IsTargetCovered}(t)\)}{%
        \(\textsc{UpdateArchive}(A, p)\)\;
        \(Z \leftarrow Z \setminus \{Z_t\}\)\;
      }{%
        \(Z_t \leftarrow Z \cup \{p\}\)\;
        \If{\(|Z_t| > N\)}{%
          \(\textsc{RemoveWorstTest}(Z_t, \delta)\)\;
        }
      }
    }
    \(\textsc{UpdateParameters}(F, R, N, M)\)\;
  }
  \Return{\(A\)}

\end{algorithm}


\subsection{Problem Representation}\label{sec:approach-representation}

As the \emph{unit} for unit test generation, we consider Python \emph{modules}.
A module is usually identical with a file
and contains definitions of,
for example,
functions, classes, or statements;
these can be nested almost arbitrarily.
When the module is loaded
the definitions and statements at the top level are executed.
While generating tests
we do not only want all definitions to be executed,
but also all structures defined by those definitions,
for example,
functions, closures, or list comprehensions.
Thus,
in order to apply a search algorithm,
we first need to define a proper representation of the valid solutions
for this problem.
We use a representation
based on prior work from the domain of testing Java code~\citep{FA13}.
For each statement~\(s_j\) in a test case~\(t_i\)
we assign one value~\(v(s_j)\)
with type~\(\tau(v(s_j)) \in \mathcal{T}\),
with the finite set of types~\(\mathcal{T}\) used in the
subject-under-test~(SUT)
and the modules transitively imported by the SUT.
A set of test cases form a \emph{test suite}.
We define five kinds of statements:
\emph{Primitive statements} represent \mintinline{python}{int},
\mintinline{python}{float}, \mintinline{python}{bool}, \mintinline{python}{bytes}
and \mintinline{python}{str} variables,
for example, \mintinline{python}{var\_0 = 42}.
Value and type of a statement are defined by the primitive variable.
Note that although in Python everything is an object,
we treat these values as primitives
because they do not require further construction in Python's syntax.
\emph{Constructor statements} create new instances of a class,
for example, \mintinline{python}{var\_0 = SomeType()}.
Value and type are defined by the constructed object;
any parameters are satisfied from the set~\(V = \{v(s_k) \mid 0 \leq k < j\}\).
\emph{Method statements} invoke methods on objects,
for example, \mintinline{python}{var\_1 = var\_0.foo()}.
Value and type are defined by the return value of the method;
source object and any parameters are satisfied from the set~\(V\)\kern-0.2em .
\emph{Function statements} invoke functions,
for example, \mintinline{python}{var\_2 = bar()}.
They do not require a source object
but are otherwise identical to method statements.
Extending our previous work~\citep{LKF20}
we introduce the generation of collection statements.
\emph{Collection statements} create new collections,
that is,
\mintinline{python}{List},
\mintinline{python}{Set},
\mintinline{python}{Dict},
and \mintinline{python}{Tuple}.
An example for such a list collection statement is
\mintinline{python}{var\_2 = [var\_0, var\_1]};
an example for a dictionary collection statement is
\verb!var_4 = {var_0: var_1, var_2: var_3}!.
Value and type are defined by the constructed collection;
elements of a collection
are satisfied from the set~\(V\)\kern-0.2em .
For dictionaries,
both keys and values are satisfied from~\(V\)\kern-0.2em .
Tuples are treated similar to lists;
their sole difference in Python is that lists are mutable
while tuples are immutable.
Previously,
we always filled in all parameters (except \mintinline{python}{*args} and \mintinline{python}{**kwargs}),
when creating a constructor, method or function statement and passed the parameters by position.
However,
filling all parameters might not be necessary, as some parameters may have default values
or are optional~(for example
\mintinline{python}{*args} and \mintinline{python}{**kwargs},
which will result in an empty tuple or dictionary, respectively).
It can also be impossible to pass certain parameters by position
as it is possible to restrict them
to be only passed by \emph{keyword}.
We improved our representation of statements with parameters,
by (1) passing parameters in the correct way,
that is,
\emph{positional} or by \emph{keyword},
and (2) leaving optional parameters empty with some probability.
Parameters of the form \texttt{*args} or \texttt{**kwargs}
capture positional or keyword arguments which are not bound to any other parameter.
Hereby, \texttt{args} and \texttt{kwargs} are just names
for the formal parameters;
they can be chosen arbitrarily,
but \texttt{args} and \texttt{kwargs} are the most common names.
Their values can be accessed as a tuple (\texttt{*args}) or a dictionary (\texttt{**kwargs}).
We fill these parameters by constructing a list or dictionary of appropriate type and passing its elements as arguments by
using the \texttt{*} or \texttt{**} unpacking operator, respectively.
Consider the example snippet in \cref{lst:test-case-examples}
to shed light into how a test case for functions involving those parameter types
may look like.
\begin{lstlisting}[%
  language=python,%
  caption={Example test cases for functions accepting lists or dictionaries},%
  label={lst:test-case-examples},%
  float=th,%
]
def my_sum(*args):
    result = 0
    for x in args:
        result += x
    return result

# Test case for my_sum
def test_case_0():
    int_0 = 42
    list_0 = [int_0, int_0]
    # Equivalent to my_sum(int_0, int_0)
    int_1 = my_sum(*list_0)
    assert int_1 == 84

def concatenate(**kwargs):
    result = ""
    for key, value in kwargs.items():
        result += key + "=" + value + ";"
    return result

# Test case for concatenate
def test_case_1():
    str_0 = 'foo'
    str_1 = 'bar'
    dict_0 = {str_0: str_1}
    # Equivalent to concatenate(foo=str_1)
    str_2 = concatenate(**dict_0)
    assert str_2 == 'foo=bar;'

\end{lstlisting}

This representation is of variable size;
we constrain the size of test cases~\(l\in[1, L]\)
and test suites~\(n\in[1, N]\).
%
In contrast to prior work on testing Java~\citep{FA13},
we do not define attribute or assignment statements;
attributes of objects are not explicitly declared in Python
but assigned dynamically,
hence it is non-trivial to identify the existing attributes of an object
and we leave it as future work.
Assignment statements in the Java representation could assign values to array indices.
This was necessary, as Java arrays can only be manipulated using the \texttt{[]}-operator.
While Python also has a \texttt{[]}-operator,
the same effect can also be achieved
by directly calling the \texttt{\_\_setitem\_\_}
or \texttt{\_\_getitem\_\_} methods.
Please note that we do not use the latter approach in \pynguin currently,
because \pynguin considers all methods
having names starting with one or more underscores to be private methods;
private methods are not part of the public interface of a module
and thus \pynguin does not directly call them.
Given these constraints,
we currently cannot generate a test case
as depicted in \Cref{lst:unsupported-statements}
because this would require some of the aforementioned features,
such as reading and writing attributes
or the \texttt{[]}-operator for list manipulation.
The latter is currently not implemented in \pynguin
because we assume
that changing the value of a list element
is not often required;
it is more important to append values to lists,
which is supported by \pynguin.
Additionally,
in Java an array has a fixed size,
whereas lists in Python have variable size.
This would require a way to find out a valid list index
that could be used for the \texttt{[]}-operator.

\begin{lstlisting}[%
  float=th,%
  caption={A test case with statements that are currently not supported},%
  label={lst:unsupported-statements},%
  language=Python,%
]
def test_case_0():
    obj_0 = SomeClass()
    # Attribute read not supported
    var_0 = obj_0.foo

    int_0 = 42
    # Attribute write not supported
    obj_0.bar = int_0

    int_1 = 0
    int_2 = 23
    list_0 = [int_0]
    # []-operator not supported.
    list_0[int_1] = int_2
\end{lstlisting}

\subsection{Test Cluster}\label{sec:approach-testcluster}
For a given subject under test, the \emph{test cluster}~\citep{WL05} defines the set of available functions and classes along with their methods and constructors.
The generation of the test cluster recursively includes all imports from other modules,
starting from the subject under test.
The test cluster also includes all primitive types
as well as Python's built-in collection types,
that are,
\mintinline{python}{List}, \mintinline{python}{Set},
\mintinline{python}{Dict}, and \mintinline{python}{Tuple}.
To create the test cluster
we load the module under test
and inspect it using the \texttt{inspect} module
from the standard Python API
to retrieve all available functions and classes
from this module.
Additionally,
we transitively inspect all dependent modules.
The resulting test cluster basically consists of two maps and one set:
the set contains information about all callable or accessible elements
in the module under test,
which are classes, functions, and methods.
This set also stores information about the fields of enums,
as well as static fields at class or module level.
During test generation,
\pynguin selects from this set the callable or accessible elements
in the module under test
to generate its inputs for.
The two maps store information about
which callable or accessible elements can generate
a specific type,
or modify it.
Please note that these two maps do not only contain elements
from the module under test
but also from the dependent modules.
\Pynguin uses these two maps
to generate or modify specific types,
if needed.

\subsection{Operators for the Genetic Algorithms}\label{sec:approach-operators}

Except the presented random algorithm~(see \cref{sec:approach-randoop}),
the test-generation algorithms~(see
\crefrange{sec:approach-ws}{sec:approach-mio})
implemented in \pynguin are genetic algorithms.
Genetic algorithms are inspired by natural evolution
and have been used to address many different optimisation problems.
They encode a solution to the problem as an individual,
called \emph{chromosome};
a set of individuals is called \emph{population}.
Using operations inspired by genetics,
the algorithm optimises the population gradually.
Operations are,
for example,
\emph{crossover}, \emph{mutation}, and \emph{selection}.
Crossover merges genetic material from at least two individuals
into a new offspring,
while mutation independently changes the elements of an individual.
Selection is being used to choose individuals for reproduction
that are considered better with respect to some fitness
criterion~\citep{CGA+18}.
In the following,
we introduce those operators
and their implementation in \pynguin
in detail.

\subsubsection{The Crossover Operator}\label{sec:approach-crossover}

Crossover is used to merge the genetic material
from at least two individuals
into a new offspring.
The different implemented genetic algorithms
use different objects as their individuals:
DynaMOSA, MIO, and MOSA consider a test case to be an individual,
whereas Whole Suite considers a full test suite,
consisting of potentially many test cases,
to be an individual.
This makes it necessary to distinguish between
\emph{test-suite crossover} and \emph{test-case crossover};
both work in a similar manner
but have subtle differences.

\paragraph{Test-suite Crossover}\label{sec:approach-testsuite-crossover}

The search operators for our representation are based on those used in \evosuite~\citep{FA13}:
We use single-point relative crossover for both, crossing over test cases and test suites.

The crossover operator for \emph{test suites},
which is used for the whole suite algorithm,
takes two parent test suites~\(P_1\) and~\(P_2\) as input,
and generates two offspring test suites~\(O_1\) and~\(O_2\),
by splitting both \(P_1\) and \(P_2\) at the same relative location,
exchanging the second parts and concatenating them with the first parts.
Individual test cases have no dependencies between each other,
thus the application of crossover
always generates valid test suites as offspring.
Furthermore,
the operator decreases the difference in the number of test cases
between the test suites,
thus \(\abs (|O_1|-|O_2|) \leq \abs(|P_1|-|P_2|)\).
Therefore,
no offspring will have more test cases than the larger of its parents.

\paragraph{Test-case Crossover}\label{sec:approach-testcase-crossover}

For the implementation of DynaMOSA, MIO, and MOSA
we also require a crossover operator for \emph{test cases}.
This operator works similar to the crossover operator for test suites.
We describe the differences in the following
using an example.
\Cref{lst:before-crossover} depicts an example
defining a class \mintinline{python}{Foo}
containing a constructor and two methods,
as well as two test cases
that construct instances of \mintinline{python}{Foo}
and execute both methods.
During crossover,
each test case is divided into two parts by the crossover point and
the latter parts of both test cases are exchanged,
which may result in the test cases depicted in \cref{lst:after-crossover}.
Since statements in the exchanged parts
may depend on variables defined in the original first part,
the statement or test case needs to be repaired to remain valid.
For example,
the insertion of the call \mintinline{python}{foo\_0.baz(int\_0)}
into the first crossed-over test case
requires an instance of \mintinline{python}{Foo}
as well as an \mintinline{python}{int} value.
In the example, the crossover operator randomly decided
to create a new \mintinline{python}{Foo} instance instead of reusing
the existing \mintinline{python}{foo\_0}
as well as creating a new \mintinline{python}{int} constant to
satisfy the \mintinline{python}{int} argument
of the \mintinline{python}{baz} method.
For the second crossed-over test case,
the operator simply reused both,
the instance of \mintinline{python}{Foo} as well the existing
\mintinline{python}{str} constant to satisfy the requirements
for the \mintinline{python}{bar} method.

\begin{lstlisting}[%
  float,%
  caption={A class under test and two generated test cases before applying crossover},%
  label={lst:before-crossover},%
  language=Python,%
  %numbers=left,%
]
# Class under test
class Foo:
    def __init__(self, foo: str):
        self._foo = foo

    def bar(self, suffix: str) -> None:
        print(self._foo + suffix)

    def baz(self, repeat: int) -> None:
        print(self._foo * repeat)

# Test cases
def test_case_0():
    str_0 = "string a"
    foo_0 = Foo(str_0)
    str_1 = "string b"
    # <- Randomly chosen crossover point
    foo_0.bar(str_1)


def test_case_1():
    str_0 = "string c"
    foo_0 = Foo(str_0)
    int_0 = 1337
    # <- Randomly chosen crossover point
    foo_0.baz(int_0)
\end{lstlisting}

\begin{lstlisting}[%
  float,%
  caption={Test cases from \cref{lst:before-crossover} after performing crossover},%
  label={lst:after-crossover},%
  language=Python,%
  %numbers=left,%
]
def test_case_after_crossover_0():
    # Statements from test_case_0
    str_0 = "string a"
    foo_0 = Foo(str_0)  # Temporarily unused after crossover
    str_1 = "string b"  # Temporarily unused after crossover
    # Statement from test_case_1 + repairing statements
    foo_1 = Foo(str_0)
    int_0 = 42
    foo_1.baz(int_0)


def test_case_after_crossover_1():
    # Statements from test_case_1
    str_0 = "string c"
    foo_0 = Foo(str_0)
    int_0 = 1337        # Temporarily unused after crossover
    # Statement from test_case_0
    foo_0.bar(str_0)    # Used str_0 to satisfy parameter
\end{lstlisting}

\subsubsection{The Mutation Operator}\label{sec:approach-mutation}

Similarly to the crossover operator,
the different granularity of the individuals
between the different genetic algorithms
requires a different handling in the mutation operation.

\paragraph{Test-suite Mutation}

When mutating a \emph{test suite}~\(T\),
each of its test cases is mutated with probability~\(\frac{1}{|T|}\).
After mutation,
we add new randomly generated test cases to~\(T\).
The first new test case is added with probability~\(\sigma_\text{testcase}\).
If it is added,
a second new test case is added with probability~\(\sigma_\text{testcase}^2\);
this happens until the \(i\)-th test case
is not added~(probability: \(1-\sigma_\text{testcase}^i\)).
Test cases are only added
if the limit~\(N\) has not been reached,
thus~\(|T| \leq N\).

\paragraph{Test-case Mutation}

The mutation of a \emph{test case}
can be one of three operations:
\emph{remove}, \emph{change}, or \emph{insert}, which we explain in the following sections.
Each of these operations can happen with the same probability
of~\(\frac{1}{3}\).
A test case that has no statements left after the application
of the mutation operator
is removed from the test suite~\(T\).

When mutating a test case \(t\) whose last execution raised an exception at statement \(s_i\), the following two rules apply in order:
\begin{enumerate}
	\item If \(t\) has reached the maximum length, that is, \(|t| \geq L\), the statements \(\{s_j\mid i < j < l\}\) are removed from \(t\).
	\item Only the statements \(\{s_j\mid 0 \leq j \leq i\}\) are considered for mutation, because the statements \(\{s_j\mid i < j < l\}\) are never reached and thus have no impact on the execution result.
\end{enumerate}
For constructing the initial population, a random test case~\(t\)
is sampled by uniformly choosing a value~\(r\) with \(1 \leq r \leq L\),
and then applying the insertion operator repeatedly starting with an empty test case~\(t'\),
until~\(|t'| \geq r\).

\paragraph{The Insertion Mutation Operation}\label{sec:approach-mutation-insert}

With probability \(\sigma_\text{statement}\)
we insert a new statement at a random position \(p \in [0,l]\).
If it is inserted,
we insert another statement with probability \(\sigma_\text{statement}^2\)
and so on, until the \(i\)-th statement is not inserted.
New statements are only inserted,
as long as the limit \(L\) has not been reached,
that is,
\(|t| < L\).
For each insertion,
with probability \(\frac{1}{2}\) each,
we either insert a new call on the module under test
or we insert a method call
on a value in the set \(\{v(s_k) \mid 0 \leq k < p\}\).
Any parameters of the selected call are either reused from the set
\(\{v(s_k) \mid 0 \leq k < p\}\),
set to \mintinline{python}{None},
possibly left empty if they are optional~(see \cref{sec:approach-representation}),
or are randomly generated.
The type of a randomly generated parameter is either defined by its type hint,
or if not available,
chosen randomly from the test cluster~(see \cref{sec:approach-testcluster}).
If the type of the parameter is defined by a type hint,
we can query the test cluster for callable elements
in the subject under test or its dependencies
that generate an object of the required type.
Generic types currently cannot be handled properly in \pynguin,
only Python's collection types are addressed.
A parameter that has no type annotation
or the annotation \texttt{Any},
requires us to consider all available types in the test cluster
as potential candidates.
For those,
we can only randomly pick an element from the test cluster.
\mintinline{python}{*args}, \mintinline{python}{**kwargs}
as well as parameters with a default value
are only filled with a certain probability.
For \mintinline{python}{*args: T} and \mintinline{python}{**kwargs: T}
we try to create or reuse a parameter of type \mintinline{python}{List[T]}
or \mintinline{python}{Dict[str, T]}, respectively.
Primitive types are either randomly initialized within a certain range
or reuse a value from static or dynamic constant seeding~\citep{FA12}
with a certain probability.
Complex types are constructed in a recursive backwards fashion,
that is,
by constructing their required parameters or reusing existing values.

\paragraph{The Change Mutation Operation}\label{sec:approach-mutation-change}

For a test case \(t = \langle s_0, s_1, \dots, s_{l-1} \rangle\)
of length \(l\),
each statement \(s_i\) is changed with probability \(\frac{1}{l}\).
For \mintinline{python}{int} and \mintinline{python}{float} primitives,
we choose a random standard normally distributed value \(\alpha\).
For \mintinline{python}{int} primitives
we add \(\alpha \Delta_\text{int}\) to the existing value.
For \mintinline{python}{float} primitives
we either add \(\alpha \Delta_\text{float}\)
or \(\alpha\) to the existing value,
or we change the amount of decimal digits of the current value
to a random value in \([0,\Delta_\text{digits}]\).
Here \(\Delta_\text{int}\),
\(\Delta_\text{float}\) and \(\Delta_\text{digits}\) are constants.
For \mintinline{python}{str} primitives,
with probability \(\frac{1}{3}\) each,
we delete, replace, and insert characters.
Each character is deleted or replaced with probability \(\frac{1}{|v(s_i)|}\).
A new character is inserted at a random location
with probability \(\sigma_\text{str}\).
If it is added,
we add another character with probability \(\sigma_\text{str}^2\) and so on,
until the \(i\)-th character is not added.
This is similar to how we add test cases to a test suite.
\mintinline{python}{bytes} primitives are mutated similar
to \mintinline{python}{str} primitives.
For \mintinline{python}{bool} primitives, we simply negate \(v(s_i)\).
For Tuples,
we replace each of its elements with probability \(\frac{1}{|v(s_i)|}\).
Lists, sets, and dictionaries are mutated similar
to how string primitives are mutated.
Values for insertion or replacement
are taken from \(\{v(s_k) \mid 0 \leq k < i\}\).
When mutating an entry of a dictionary,
with probability \(\frac{1}{2}\) we either replace the key or the value.
For method, function, and constructor statements,
we change each argument of a parameter with probability \(\frac{1}{p}\),
where \(p\) denotes the number of formal parameters of the callable
used in \(s_i\).
For methods, this also includes the callee.
If an argument is changed
and the parameter is considered optional~(see
\cref{sec:approach-representation})
then with a certain probability the associated argument is removed,
if it was previously set,
or set with a value from \(\{v(s_k) \mid 0 \leq k < i\}\) if it was not set.
Otherwise, we either replace the argument
with a value from \(\{v(s_k) \mid 0 \leq k < i\}\),
whose type matches the type of the parameter or use \texttt{None}.
If no argument was replaced,
we replace the whole statement \(s_i\)
by a call to another method, function, or constructor,
which is randomly chosen from the test cluster,
has the same return type as \(v(s_i)\),
and whose parameters can be satisfied
with values from \(\{v(s_k) \mid 0 \leq k < i\}\).

\paragraph{The Remove Mutation Operation}\label{sec:approach-mutation-remove}

For a test case \(t = \langle s_0, s_1, \dots, s_{l-1} \rangle\)
of length \(l\),
each statement \(s_i\) is deleted with probability \(\frac{1}{l}\).
As the value \(v(s_i)\) might be used as a parameter
in any of the statements \(s_{i+1},\dots, s_{l-1}\),
the test case needs to be repaired in order to remain valid.
For each statement \(s_j\), \(i < j < l\),
if \(s_j\) refers to \(v(s_i)\),
then this reference is replaced with another value
out of the set \(\{v(s_k) \mid 0 \leq k < j \land k \neq i \}\),
which has the same type as \(v(s_i)\).
If this is not possible, then \(s_j\) is deleted as well recursively.
%


\subsection{Covering and Tracing Python Code}\label{sec:approach-covering}

A Python module contains various control structures,
for example,
\mintinline{python}{if} or \mintinline{python}{while} statements,
which are guarded by logical predicates.
The control structures are represented by conditional jumps
at the bytecode level,
based on either a unary or binary predicate.
We focus on \emph{branch coverage} in this work,
which requires that each of those predicates evaluates to
both true and false.
Let~\(B\) denote the set of branches in the subject under test---two
for each conditional jump in the byte code.
Everything executable in Python is represented as a \emph{code object}.
For example,
an entire module is represented as a code object,
a function within that module is represented as another code object.
We want to execute all code objects~\(C\) of the subject under test.
Therefore,
we keep track of the executed code objects~\(C_T\)
as well as the minimum \emph{branch distance}~\(d_{\min}(b, T)\)
for each branch~\(b \in B\),
when executing a test suite~\(T\).
\(B_T \subseteq B\) denotes the set of taken branches.
Code objects which contain branches do not have to be considered as individual coverage targets,
since covering one of their branches also covers the respective code object.
Thus, we only consider the set of branch-less code objects~\(C_{L} \subseteq
C\).
We then define the branch coverage~\(\cov(T)\) of a test suite~\(T\) as
\(  \cov(T) = \frac{|C_T \cap C_{L}| + |B_T|}{|C_{L}| + |B|} \).

\emph{Branch distance} is a heuristic to determine
how far a predicate is away from evaluating to true or false, respectively.
In contrast to previous work on Java,
where most predicates at the bytecode level operate only
on Boolean or numeric values,
in our case the operands of a predicate can be any Python object.
Thus,
as noted by \citet{A13},
we have to define our branch distance in such a way
that it can handle arbitrary Python objects.
Let~\(\mathbb{O}\) be the set of possible Python objects
and let~\(\Theta := \{\equiv, \not\equiv, <, \leq, >, \geq, \in, \notin,
=, \neq\}\) be the set of binary comparison operators~(remark: we use
\enquote{\(\equiv\)}, \enquote{\(=\)}, and \enquote{\(\in\)} for Python's
\mintinline{python}{==}, \mintinline{python}{is}, and \mintinline{python}{in}
keywords, respectively).
For each~\(\theta \in \Theta\),
we define a function~\(\delta_\theta: \mathbb{O}\times\mathbb{O} \to
\mathbb{R}_0^+ \cup \{\infty\}\)
that computes the branch distance of the true branch of a predicate of the form
\(a \mathbin{\theta} b\), with \(a,b\in\mathbb{O}\) and \(\theta\in\Theta\).
By~\(\delta_{\bar \theta}(a,b)\) we denote the distance of the false branch,
where~\(\bar \theta\) is the complementary operator of~\(\theta\).
Let further \(k\) be a positive number,
and let~\(\lev(x,y)\) denote the Levenshtein distance~\citep{L66}
between two strings~\(x\) and~\(y\).
The value of \(k\) is used in cases where we know that the distance is not \(0\),
but we cannot compute an actual distance value,
for example,
when a predicate compares two references for identity,
the branch distance of the true branch is either \(0\),
if the references point to the same object,
or \(k\),
if they do not.
While the actual value of \(k\) does not matter, we use \(k = 1\).

The predicates \(\mathrm{is\_numeric}(z)\), \(\mathrm{is\_string}(z)\)
and \(\mathrm{is\_bytes}(z)\)
determine whether the type of their argument~\(z\) is numeric, a string
or a byte array, respectively.
The function \(\mathrm{decode}(z)\) decodes a byte array into a string
by decoding every byte into a unique character, for example, by using
the encoding ISO-8859-1.

\begin{align*}
\delta_{\equiv}(a, b) &= \begin{cases}
0 & a \equiv b \\
|a-b| & a \not\equiv b \land \text{is\_numeric}(a) \land \text{is\_numeric}(b) \\
\text{lev}(a,b) & a \not\equiv b \land \text{is\_string}(a) \land \text{is\_string}(b) \\
\text{lev}(\text{decode}(a),\text{decode}(b)) & a \not\equiv b \land \text{is\_bytes}(a) \land \text{is\_bytes}(b) \\
\infty & \text{otherwise}
\end{cases}
\\
\delta_{<}(a, b) &= \begin{cases}
0 & a < b \\
a-b+k & a \geq b \land \text{is\_numeric}(a) \land \text{is\_numeric}(b) \\
\infty & \text{otherwise}
\end{cases}
\\
\delta_{\leq}(a, b) &= \begin{cases}
0 & a \leq b \\
a-b+k & a > b \land \text{is\_numeric}(a) \land \text{is\_numeric}(b) \\
\infty & \text{otherwise}
\end{cases}
\\
\delta_{>}(a, b) &= \delta_{<}(b,a)
\\
\delta_{\geq}(a, b) &= \delta_{\leq}(b,a)
\\
\delta_{\in}(a, b) &= \begin{cases}
	0 & a \mathbin{\in} b \\
	\min(\{\delta{\equiv}(a,x) \mid x \in b \} \cup \{\infty\}) & \text{otherwise}
\end{cases}
\\
\delta_{\theta}(a, b) &= \begin{cases}
0 & a \mathbin{\theta} b \\
k & \text{otherwise}
\end{cases}
\qquad \theta \in \{\not\equiv, \notin, =, \neq\}
\end{align*}
Note that every object in Python represents a Boolean value
and can therefore be used as a predicate.
Classes can define how their instances are coerced into a Boolean value,
for example, numbers representing the value zero or empty collections are interpreted as false,
whereas non-zero numbers or non-empty collections are interpreted as true.
We assign a distance of~\(k\) to the true branch,
if such a unary predicate \(v\) represents false.
Otherwise, we assign a distance of \(\delta_F(v)\) to the false branch,
where \(\mathrm{is\_sized}(z)\) is a predicate that determines if its argument \(z\) has a size
and \(\mathrm{len}(z)\) is a function that computes the size of its argument \(z\).
\begin{align*}
\delta_F(a) &= \begin{cases}
\text{len}(a) & \text{is\_sized}(a)\\
|a| & \text{is\_numeric}(a)\\
\infty & \text{otherwise}
\end{cases}
\end{align*}

Future work shall refine the branch distance for different operators
and operand types.

\subsection{Fitness Functions}\label{sec:approach-fitness}

The fitness function required by genetic algorithms
is an estimate of how close an individual
is towards fulfilling a specific goal.
As stated before
we optimise our generated test suites towards maximum branch coverage.
We define our fitness function
with respect to this coverage criterion.
Again,
we need to distinguish between the fitness function for Whole Suite,
which operates on the test-suite level,
and the fitness function for DynaMOSA, MIO, and MOSA,
which operates on the test-case level.

\subsubsection{Test-suite Fitness}\label{sec:approach-testsuite-fitness}

The fitness function required by our Whole Suite approach
is constructed similar to the one used in \toolname{EvoSuite}~\citep{FA13}
by incorporating the branch distance.
The fitness function estimates how close a test suite is
to covering \emph{all} branches of the SUT.
Thus,
every predicate has to be executed at least twice,
which we enforce in the same way as existing work~\citep{FA13}:
the actual branch distance~\(d(b, T)\) is given by
\begin{align*}
d(b,T) &= \begin{cases}
0 & \text{if the branch has been covered}\\
\nu(d_{min}(b,T)) & \text{if the predicate has been executed at least twice}\\
1 & \text{otherwise}
\end{cases}
\end{align*}
with \(\nu(x) = \frac{x}{x+1}\) being a normalisation function~\citep{FA13}.
Note that the requirement of a predicate being executed at least twice
is used to avoid a circular behaviour
during the search, where the
whole suite algorithm alternates between covering a branch of a predicate,
but in doing so loses coverage on the opposing branch~\citep{FA13}.

Finally,
we can define the resulting fitness function~\(f\) of a test suite~\(T\) as
\begin{align*}
  f(T) &= |C_{L} \setminus C_T| + \sum_{b \in B} d(b, T)
\end{align*}

\subsubsection{Test-case Fitness}\label{sec:approach-testcase-fitness}

The aforementioned fitness function is only applicable
for the Whole Suite algorithm,
which considers every coverage goal at the same time.
DynaMOSA, MIO, and MOSA
consider individual coverage goals.
Such a goal is to cover either a certain branch-less code object~\(c \in C_{L}\)
or a branch~\(b \in B\).
For the former, its fitness is given by
\begin{align*}
  f_c(t) &= \begin{cases}
    0 & \text{\(c\) has been executed}\\
    1 & \text{otherwise}
  \end{cases}
\end{align*}
while for the latter it is defined as
\begin{align*}
f_b(t) = al(b,t) + \nu(d_{min}(b, t))
\end{align*}
where \(d_{min}(b, t)\) is the smallest recorded branch distance of branch~\(b\)
when executing test case \(t\)
and \(al(b, t)\) denotes the approach level.
The \emph{approach level} is the number of control dependencies
between the closest executed branch and~\(b\)~\citep{PKT18}.

\subsection{Assertion Generation}\label{sec:approach-assertion-generation}

\Pynguin aims to generate \emph{regression} assertions~\citep{Xie06}
within its generated test cases
based on the values it observes during test-case execution.
We consider the following types as \emph{assertable}:
enum values,
builtin types (\texttt{int}, \texttt{str}, \texttt{bytes},
\texttt{bool}, and \texttt{None}),
as well as builtin collection types (\texttt{tuple}, \texttt{list},
\texttt{dict}, and \texttt{set})
as long as they only consist of assertable elements.
Additionally,
\pynguin can generate assertions for \texttt{float} values
using an approximate comparison
to mitigate the issue of imprecise number representation.
For those assertable types,
\pynguin can generate the assertions directly,
by creating an appropriate expected value
for the assertion's comparison.
We call other types non-assertable.
This does,
however,
not mean that \pynguin is not able to generate assertions for these types.
It indicates that \pynguin will not try to create a specific object
of such a type
as an expected value for the assertion's comparison,
but it will aim to check the object's state
by inspecting its public attributes;
assertions shall then be generated on the values
of these attributes, if possible.
For non-assertable collections,
\pynguin creates an assertion on their size
by calling the \texttt{len} function on the collection
and assert for that size.
As a fallback,
if none of the aforementioned strategies is successful,
\pynguin is asserting that an object is not \texttt{None}.
\Cref{lst:example-assertions-regular} shows a simplified example
of how test cases for the code snippet in \cref{lst:example-assertion-code}
could look like.
Please note
that the example was not directly generated with \pynguin
but manually simplified for demonstration purpose.

\begin{lstlisting}[%
  float=th,
  caption={A code snippet to generate tests with assertions for.},
  label={lst:example-assertion-code},
  language=Python,
]
import queue

def divide(x: int, y: int) -> float:
  if y < 0:
    raise ValueError("y shall only be positive")
  return x / y

def queue():
  return queue.Queue()
\end{lstlisting}

\begin{lstlisting}[%
  float=th,
  caption={Some test cases for the snippet in
      \cref{lst:example-assertion-code}.},
  label={lst:example-assertions-regular},
  language=Python,
]
import example as module0
import pytest

def test_case_0:
  int_0 = 1
  int_1 = 2
  float_0 = module0.divide(int_0, int_1)
  assert pytest.approx(0.5) == float_0

def test_case_1:
  var_0 = module_0.queue()
  assert var_0 is not None
\end{lstlisting}

Additionally,
\pynguin creates assertions on raised exceptions.
Python does,
in contrast to Java,
not know about the concept of checked and unchecked,
or expected and unexpected exceptions.
During \pynguin's analysis of the module under test
we analyse the abstract syntax tree of the module
and aim to extract for each function
which exceptions it explicitly raises but not catches
or that are specified in the function's Docstring.
We consider these exceptions to be \emph{expected},
all other exceptions that might appear during execution
are \emph{unexpected}.
For an expected exception
a generated test case
will utilise the \texttt{pytest.raises} context
from the PyTest framework\footnote{
  \url{https://www.pytest.org}, accessed 2022–07–13.%
},
which checks whether an exception of a given type
was raised during execution
and causes a failure of the test case if not.
\Pynguin decorates a test case,
who's execution causes an unexpected exception,
with the \texttt{pytest.mark.xfail} decorator
from the PyTest library;
this decorator marks the test case as expected to be failing.
Since \pynguin is not able to distinguish
whether this failure is expected
it is up to the user of \pynguin to inspect
and deal with such a test case.
\Cref{lst:example-assertion-exceptions} shows two test cases
that check for exceptions.

\begin{lstlisting}[%
  float=th,%
  caption={Some test cases handling exceptions for the code snippet
      in \cref{lst:example-assertion-code}.},%
  label={lst:example-assertion-exceptions},%
  language=Python,%
]
import example as module0
import pytest

def test_case_0:
  int_0 = 1
  int_1 = -1
  with pytest.raises(ValueError):
    module0.divide(int_0, int_1)

@pytest.mark.xfail(strict=True)
def test_case_0:
  int_0 = 1
  int_1 = 0
  module0.divide(int_0, int_1)
\end{lstlisting}

To generate assertions,
\pynguin observes certain properties of the module under test.
The parts of interest are the set of return values
of function or method invocations and constructor calls,
the static fields on used modules,
and the static fields on seen return types.
It does so after the execution of each statement of a test case.
First,
\pynguin executes every test case twice in random order.
We do this to remove trivially flaky assertions,
for example,
assertions based on memory addresses or random numbers.
Only assertions on values that are the same on both executions
are kept as the \emph{base assertions}.
To minimise the set of assertions~\citep{FZ12}
to those that are likely to be relevant
to ensure the quality of the current test case,
\pynguin utilises a customised fork of \toolname{MutPy}\footnote{%
  \url{https://github.com/se2p/mutpy-pynguin}, accessed 2022–06–12.%
} to generate mutants of the module under test.
A mutant is a variant of the original module under test
that differs by few syntactic changes~\citep{ABD+79,DLS78,JH11}.
There exists a wide selection of so called \emph{mutation operators}
that generate mutants from a given program.
From the previous executions,
\pynguin knows the result of each test case.
It then executes all test cases against all mutants.
If the outcome for one test case differs on a particular mutant
between the original module and that mutant,
the mutant is said to be \emph{killed};
otherwise it is classified as \emph{survived}.
Out of the base assertions,
only those assertions are considered relevant
that have an impact on the result of the test execution.
Such an impact means that the test is able to kill the mutant
based on the provided assertions.
Assertions that have not become relevant
throughout all runs
are removed from the test case,
as they do not seem to help for fault finding.
%


\subsection{Limitations of \pynguin}\label{sec:approach-limitations}

At its current state,
\pynguin is still a research prototype facing several technical limitations
and challenges.
First,
we noticed that \pynguin does not isolate the test executions
properly in all cases.
We used Docker containers to prevent \pynguin from causing harm on our systems;
however,
a proper isolation of file-system access
or calls to specific functions,
such as \texttt{sys.exit},
need to be handled differently.
One of the reasons
\pynguin fails on target modules is
that they call \texttt{sys.exit} in their code,
which stops the Python interpreter
and thus also \pynguin.
Future work needs to address those challenges by,
for example,
providing a virtualised file system to \pynguin executions
or mocking calls to critical functions,
such as \texttt{sys.exit}.
%

The dynamic nature of Python allows developers
to change many properties of the program during runtime.
One example is a non-uniform object layout;
instances of the same class may have different methods or fields
during runtime,
of which we do not know when we analyse the class
before execution.
This is due to the fact that attributes can be added, removed, or changed
during runtime,
special methods like \texttt{\_\_getattr\_\_} can fake non-existing attributes
and many more.
The work of \citet{CML+18},
for example,
lists some of these dynamic features of the Python language
and their impact on fixing bugs.
%

Besides the dynamic features of the programming language
\pynguin suffers from supporting all language features.
Constructs such as generators,
iterators,
decorators,
coroutines,
or higher-order functions provide open challenges for \pynguin
and aim for future research.
Furthermore,
Python allows to implement modules in C/C++,
which are then compiled to native binaries,
in order to speed up execution.
The native binaries,
however,
lack large parts of information
that the Python source files provide,
for example,
there is no abstract syntax tree for such a module that could be analysed.
\Pynguin is not able to generate tests for a binary module
because its coverage measurement
relies on the available Python interpreter byte code
that we instrument to measure coverage.
Such an on-the-fly instrumentation is obviously not possible
for binaries in native code.
%

Lastly,
we want to note that the type system of Python
is challenging.
Consider the two types \texttt{List[Tuple[Any, ...]}
and \texttt{List[Tuple[int, str]]};
there is no built-in support in the language to decide
at runtime whether one is a subtype of the other.
Huge engineering effort has been put into static type checkers,
such as \texttt{mypy}\footnote{%
  \url{http://mypy-lang.org/}, accessed 2022–07–14.%
},
however they are not considered feature complete.
Additionally,
each new version of Python brings new features regarding type information
to better support developers.
This,
however,
also causes new challenges for tool developers to support those new features.

\section{Empirical Evaluation}\label{sec:evaluation}

We use our \pynguin test-generation framework
to empirically study automated unit test generation in Python.
A crucial metric to measure the quality of a test suite
is the coverage value it achieves;
a test suite cannot reveal any faults
in parts of the subject under test~(SUT)
that are not executed at all.
Therefore,
achieving high coverage
is an essential property
of a good test suite.
%

The characteristics of different test-generation approaches
have been extensively studied
in the context of statically typed languages,
for example by \citet{PKT18a} or \citet{CGA+18}.
With this work we aim to determine
whether these previous findings on the performance of test generation techniques
also generalise to Python with respect to coverage:

\begin{resq}[RQ\ref{rq:quality}]\label{rq:quality}
  How do the test-generation approaches compare on Python code?
\end{resq}

Our previous work~\citep{LKF20} indicated
that the availability of type information
can be crucial to achieve higher coverage values.
Type information,
however,
is not available for many programs written in dynamically typed languages.
To gain information on the influence of type information
on the different test-generation approaches,
we extend our experiments from our previous study~\citep{LKF20}.
Thus,
we investigate the following research question:

\begin{resq}[RQ\ref{rq:typeinfluence}]\label{rq:typeinfluence}
  How does the availability of type information influence test generation?
\end{resq}

Coverage is one central building block of a good test suite.
Still,
a test-generation tool solely focussing on coverage
is not able to create a test suite
that reveals many types of faults from the subject under test.
The effectiveness of the generated assertions
is therefore a crucial metric
to reveal faults.
We aim to investigate the quality of the generated assertions by asking:

\begin{resq}[RQ\ref{rq:assertions}]\label{rq:assertions}
  How effective are the generated assertions at revealing faults?
\end{resq}


\subsection{Experimental Setup}\label{sec:evaluation-setup}

The following sections describe the setup
that we used for our experiments.


\subsubsection{Experiment Subjects}\label{sec:evaluation-subjects}

For our experiments
we created a dataset of Python projects,
which allows us to answer our research questions.
We only selected projects for the dataset
that contain type information in the style of PEP~484.
Projects without type information
would not allow us to investigate
the influence of the type information~(see RQ\ref{rq:typeinfluence}).
Additionally,
we focus on projects that use features of the Python programming language
that are supported by \pynguin.
We do this to avoid skewing the results
due to missing support for such features in \pynguin.
%

We excluded projects with dependencies to native code
such as \texttt{numpy} from our dataset.
We do this for two reasons:
first,
we want to avoid compiling these dependencies,
which would require a C/C++ compilation environment
in our experiment environment
to keep this environment as small as possible.
Second,
\pynguin's module analysis and instrumentation
rely on the source code
and Python's internally used byte code.
Both are not available for native code,
which can limit \pynguin's applicability on such projects.
Unfortunately,
many projects depend to native-code libraries
such as \texttt{numpy},
either directly or transitively.
\Pynguin can still be used on such projects
but its capabilities might be limited
by not being able to analyse these native-code dependencies.
Fortunately,
only few Python libraries are native-code only.
Dependencies that only exist in native code,
for example,
the widely used \texttt{numpy} library,
can still be used,
although \pynguin might not be able
to retrieve all information it could retrieve from a Python source file.
%

We reuse nine of the ten projects from our previous work~\citep{LKF20};
the removed project is \texttt{async\_btree},
which provides an asynchronous behaviour-tree implementation.
The functions in this project are implemented as co-routines,
which would require a proper handling of the induced parallelism
and special function invocations.
\Pynguin currently only targets sequential code;
it therefore cannot deal with this form of parallelism
to avoid any unintended side effects.
Thus,
for the \texttt{async\_btree} project,
the test generation fails
and only import coverage can be achieved,
independent of the used configuration.
None of the other used projects relies on co-routines.
%

We selected two projects from the \textsc{BugsInPy} dataset~\citep{WSL+20},
a collection of 17 Python projects at the time of writing.
The selected projects are those projects in the dataset
that provide type annotations.
The remaining nine projects have been randomly selected
from the \toolname{ManyTypes4Py} dataset~\citep{MLG21},
a dataset of more than \num{5200} Python repositories
for evaluating machine learning-based type inference.
We do not use all projects from \toolname{ManyTypes4Py}
for various reasons:
first, generating tests for \num{183916} source code files
from the dataset
would cause a tremendous computational effort.
Second,
many of those projects also have dependencies
on native-code libraries,
which we explicitly excluded from our selection.
Third,
the selection criterion for projects in that dataset was
that they use the \toolname{MyPy} type checker as a project dependency
in some way.
The authors state that only \SI{27.6}{\percent} of the source-code files
have type annotations.
We manually inspected the projects from \toolname{ManyTypes4Py}
we incorporated into our dataset
to ensure that they have type annotations present.
%

Overall, this results in a set of \num{\numTotalClasses} modules
from \num[round-precision=2]{\numProjects} projects.
\Cref{tab:projects} provides a more detailed overview of the projects.
The column \emph{Project Name} gives the name of the project on PyPI;
the lines of code were measured
using the \textsc{cloc}\footnote{
  \url{https://github.com/AlDanial/cloc},
  accessed 2022–07–06.%
} tool.
The table furthermore shows the total number of code objects
and predicates in a project's modules,
as well as the number of types detected per project.
The former two measures give an insight on the project's complexity:
higher numbers indicate larger complexity.
The latter provides an overview how many types \pynguin was able to parse.
Please note that \pynguin may not be able to resolve all types
as it relies on Python's \texttt{inspect} library for this task.
This library is the standard way
of inspecting modules in Python—to extract
information about existing classes and functions,
or type information.
If \texttt{inspect} fails to resolve a type
that type will not be known to \pynguin's test cluster,
which means that \pynguin will not try to generate
an object of such a type.
However,
since the \texttt{inspect} library is part of the standard library
we would consider its quality to be very good;
furthermore,
if such a type is reached in a transitive dependency,
it might be used anyway.

\begin{table}[t]
  \centering
  \tabcaption{\label{tab:projects}Projects used for evaluation}
  \begin{adjustbox}{max width=\textwidth}
    \begin{tabular}{@{} l r r r r r r @{}}
\toprule
Project Name & Version & {LOCs} & {Modules} & {CodeObjs.} & {Preds.} & {Types} \\ \midrule
\verb|apimd| & 1.2.1 & 390 & 1 & 50 & 110 & 10\\
\verb|codetiming| & 1.3.0 & 89 & 2 & 7 & 5 & 4\\
\verb|dataclasses-json| & 0.5.2 & 891 & 5 & 75 & 76 & 23\\
\verb|docstring_parser| & 0.7.3 & 506 & 4 & 50 & 86 & 9\\
\verb|flake8| & 3.9.0 & 542 & 8 & 80 & 41 & 0\\
\verb|flutes| & 0.3.0 & 235 & 3 & 4 & 0 & 8\\
\verb|flutils| & 0.7 & 772 & 6 & 48 & 136 & 19\\
\verb|httpie| & 2.4.0 & 1274 & 18 & 152 & 129 & 22\\
\verb|isort| & 5.8.0 & 841 & 6 & 37 & 9 & 11\\
\verb|mimesis| & 4.1.3 & 685 & 21 & 150 & 55 & 4\\
\verb|pdir2| & 0.3.2 & 461 & 5 & 43 & 46 & 9\\
\verb|py-backwards| & 0.7 & 618 & 18 & 125 & 100 & 13\\
\verb|pyMonet| & 0.12.0 & 394 & 8 & 145 & 59 & 6\\
\verb|pypara| & 0.0.24 & 877 & 7 & 232 & 70 & 15\\
\verb|python-string-utils| & 1.0.0 & 421 & 3 & 76 & 101 & 7\\
\verb|pytutils| & 0.4.1 & 943 & 19 & 62 & 80 & 4\\
\verb|sanic| & 21.3.2 & 1886 & 18 & 143 & 187 & 30\\
\verb|sty| & 1.0.0rc1 & 136 & 3 & 18 & 4 & 2\\
\verb|thonny| & 3.3.6 & 794 & 5 & 66 & 232 & 1\\
\verb|typesystem| & 0.2.4 & 157 & 3 & 23 & 19 & 13\\
\midrule
Total & & 12912 & 163 & 1586 & 1545 & 210 \\ \bottomrule
\end{tabular}

  \end{adjustbox}
\end{table}


\subsubsection{Experiment Settings}\label{sec:evaluation-settings}

We executed \pynguin on each of the constituent modules in sequence
to generate test cases.
We used \pynguin in version 0.25.2~\citep{Pynguin-0.25.2} in a Docker container.
The container is used for process isolation
and is based on Debian~10 together with Python~3.10.5~(the
\texttt{python:3.10.5-slim-buster} image from Docker Hub
is the basis of this container).
%

We ran \pynguin in ten different configurations:
the five generation approaches \emph{DynaMOSA}, \emph{MIO},
\emph{MOSA}, \emph{Random}, and \emph{WS}~(see
\cref{sec:approach-algorithms} for a description of each approach),
each with and without incorporated type information.
We will refer to a configuration with incorporated type information
by adding the suffix \emph{TypeHints} to its name,
for example,
\emph{DynaMOSA-TypeHints}
for DynaMOSA with incorporated type information;
the suffix \emph{NoTypes} indicates a configuration
that omits type information,
for example,
\emph{Random-NoTypes}
for feedback-directed random test generation without type information.
%

We adopt several default parameter values from \evosuite~\citep{FA13}.
It has been empirically shown~\citep{AF13}
that these default values give reasonably acceptable results.
We leave a detailed investigation of the influence
of the parameter values as future work.
The following values are only set if applicable for a certain configuration.
For our experiments,
we set the population size to \num[round-precision=2]{50}.
For the MIO algorithm,
we cut the search budget to half for its exploration phase
and the other half for its exploitation phase.
In the exploration phase,
we set the number of tests per target to ten,
the probability to pick either a random test
or from the archive to \SI{50}{\percent},
and the number of mutations to one.
For the exploitation phase,
we set the number of tests per target to one,
the probability to pick either a random test
or from the archive to zero,
and the number of mutations to ten.
We use a single-point crossover
with the crossover probability set to \num{0.750}.
Test cases and test suite are mutated using uniform mutation
with a mutation probability of \(\frac{1}{l}\),
where \(l\) denotes the number of statements
contained in the test case.
\pynguin uses tournament selection
with a default tournament size of \num[round-precision=1]{5}.
The search budget is set to \SI{600}{\second}.
This time does not include preprocessing and pre-analysis of the subject.
Neither does it include any post-processing,
such as assertion generation
or writing test cases to a file.
The search can stop early if \SI{100}{\percent} coverage is achieved
before consuming the full search budget.
Please note that the stopping condition for the search budget
is only checked between algorithm iterations.
Thus,
it may be possible that some executions of \pynguin
slightly exceed the \SI{600}{\second} search budget
before they stop the search.
%

To minimise the influence of randomness
we ran \pynguin \num[round-precision=2]{\numIterations} times
for each configuration and module.
All experiments were conducted on dedicated compute servers
each equipped with an AMD EPYC 7443P CPU and \SI{256}{\giga\byte} RAM,
running Debian~10.
We assigned each run one CPU core
and four gigabytes of RAM.
%


\subsubsection{Evaluation Metrics}\label{sec:evaluation-metrics}

We use code coverage
to evaluate the performance of the test generation
in RQ\ref{rq:quality} and RQ\ref{rq:typeinfluence}.
In particular,
we measure branch coverage at the level of Python bytecode.
Branch coverage is defined as the number of covered,
that is,
executed branches in the subject under test
divided by the total number of branches in the subject under test.
Similar to Java bytecode,
complex conditions are compiled to nested branches with atomic conditions
in Python code.
We also keep track of coverage over time
to shed light on the speed of convergence of the test-generation process;
besides,
we note the final overall coverage.
%

To evaluate the quality of the generated assertions in RQ\ref{rq:assertions}
we compute the mutation score.
We use a customised version of the \toolname{MutPy} tool~\citep{DH14}
to generate the mutants from the original subject under test.
\toolname{MutPy} brings a large selection of standard mutation operators
but also operators that are specific to Python.
A mutant is a variant of the original subject under test
obtained by injecting artificial modifications into them.
It is referred to as \emph{killed}
if there exists a test that passes on the original subject under test
but fails on the mutated subject under test.
The mutation score is defined as the number of killed mutants
divided by the total number of generated mutants~\citep{JH11}.
%

We statistically analyse the results to see
whether the differences between two different algorithms or configurations
are statistically significant or not.
We exclude modules from the further analysis
for which \pynguin failed to generate tests
\num[round-precision=2]{\numIterations} times in all respective configurations
to make the configurations comparable.
We use the non-parametric Mann-Whitney \(U\)-test~\citep{MW47}
with a \pvalue{} threshold \num{0.05} for this.
A \pvalue below this threshold indicates that the null hypothesis
can be rejected in favour of the alternative hypothesis.
In terms of coverage,
the null hypothesis states that none of the compared algorithms
reaches significantly higher coverage;
the alternative hypothesis states that one of the algorithms
reaches significantly higher coverage values.
We use the Vargha and Delaney effect size \effectsize~\citep{VD00}
in addition to testing for the null hypothesis.
The effect size states the magnitude of the difference
between the coverage values
achieved by two different configurations.
For equivalent coverage
the Vargha-Delaney statistics is \(\effectsize = \num{0.5}\).
When comparing two configurations~\(C_1\) and~\(C_2\) in terms of coverage,
an effect size of \(\effectsize > \num{0.5}\) indicates
that configuration~\(C_1\) yields higher coverage than~\(C_2\);
vice versa for \(\effectsize < \num{0.5}\).
Furthermore,
we use Pearson's correlation coefficient~\(r\)~\citep{Pea95}
to measure the linear correlation between two sets of data.
We call a value of \(r = \pm \num{1}\) a \emph{perfect correlation},
a value of~\(r\) between~\(\pm\num{0.5}\) and~\(\pm \num{1}\)
a \emph{strong correlation},
a value of~\(r\) between~\(\pm\num{0.3}\) and~\(\pm\num{0.499}\)
a \emph{medium correlation},
and a value of~\(r\) between~\(\pm\num{0.1}\) and~\(\pm\num{0.299}\)
a \emph{small} or \emph{weak correlation}.
%

Finally,
please note that we report all numbers
during our experiments
rounded to three significant digits,
except if they are countable,
such as,
for example,
lines of code in a module.
%


\subsubsection{Data Availability}\label{sec:evaluation-data}

We make all used projects and tools
as well as the experiment and data-analysis infrastructure
together with the raw data
available on Zenodo~\citep{Luk22}.
This shall allow further analysis and replication of our results.
%


\subsection{Threats to Validity}\label{sec:evaluation-threats}

As usual, our experiments are subject to a number of threats to validity.

%


\subsubsection{Internal Validity}\label{sec:evaluation-threats-internal}

The standard coverage tool for Python is \toolname{Coverage.py},
which offers the capability to measure both line and branch coverage.
It,
however,
measures branch coverage by comparing the transitions between sources lines
that have occurred and that are possible.
Measuring branch coverage using this technique
is possibly imprecise.
Not every branching statement necessarily leads to a source line transition,
for example,
\texttt{x = 0 if y > 42 else 1337} fits on one line
but contains two branches,
which are not considered by \toolname{Coverage.py}.
We thus implemented our own coverage measurement
based on bytecode instrumentation.
By providing sufficient unit tests for it we try to mitigate possible errors
in our implementation.
%

Similar threats are introduced by the mutation-score computation.
A selection of mutation-testing tools for Python exist,
however,
each has some individual drawbacks,
which make them unsuitable for our choice.
Therefore,
we implemented the computation of mutation scores ourselves.
However,
the mutation of the subject under test itself is done
using a customised version of the \toolname{MutPy}
mutation testing tool~\citep{DH14}
to better control this threat.
%

A further threat to the internal validity
comes from probably flaky tests;
a test is flaky when its verdict changes non-deterministically.
Flakiness is reported to be a problem,
not only for Python test suites~\citep{GLK+21}
but also for automatically generated tests in general~\citep{Fan19,PKH+22}.
%

The used Python inspection to generate the test
cluster~(see~\cref{sec:approach-testcluster})
cannot handle types provided by native dependencies.
We mitigate this threat
by excluding projects that have dependencies with native code.
This,
however,
does not exclude any functions from the Python standard library,
which is partially also implemented in C,
and which could influence our results.
%


\subsubsection{External Validity}\label{sec:evaluation-threats-external}

We used \num{\numTotalClasses} modules from different Python projects
for our experiments.
It is conceivable that the exclusion of projects without type annotations
or native-code libraries leads to a selection of smaller projects,
and the results may thus not generalise to other Python projects.
Furthermore,
to make the different configurations comparable,
we omitted all modules from the final evaluation
for that \pynguin was not able to generate test cases for each configuration
and each of the \num[round-precision=2]{\numIterations} iterations.
This leads to \num[round-precision=3]{\numUniqueClasses} modules
for RQ\ref{rq:quality} and RQ\ref{rq:typeinfluence}
and \num[round-precision=3]{\numUniqueClassesWithAssertions} modules
for RQ\ref{rq:assertions}.
The number of used modules for RQ\ref{rq:assertions} is lower
because we exclude modules from the analysis
that did not yield \num[round-precision=2]{\numIterations} results.
Reasons for such failures are,
for example,
flaky tests.
However,
besides the listed constraints,
no others were applied during this selection.
%


\subsubsection{Construct Validity}\label{sec:evaluation-threats-construct}

Methods called with wrong input types
still may cover parts of the code
before possibly raising exceptions
due to the invalid inputs.
We conservatively included all coverage in our analysis,
which may improve coverage for configurations
that ignore type information.
A configuration that does not use type information
will randomly pick types
to generate argument values,
although these types might be wrong.
In contrast,
configurations including type information
will attempt to generate the correct type;
they will only use a random type with small probability.
Thus,
this conservative inclusion might reduce the effect we observed.
It does,
however,
not affect our general conclusions.
%

Additionally,
we have not applied any parameter tuning to the search parameters
but use default values,
which have been shown to be reasonable choices in practice~\citep{AF13}.
%


\subsection{RQ\ref{rq:quality}: Comparison of the Test-Generation Approaches}%
\label{sec:evaluation-rq1}

\begin{figure}[th]
  \centering
  \includegraphics[]{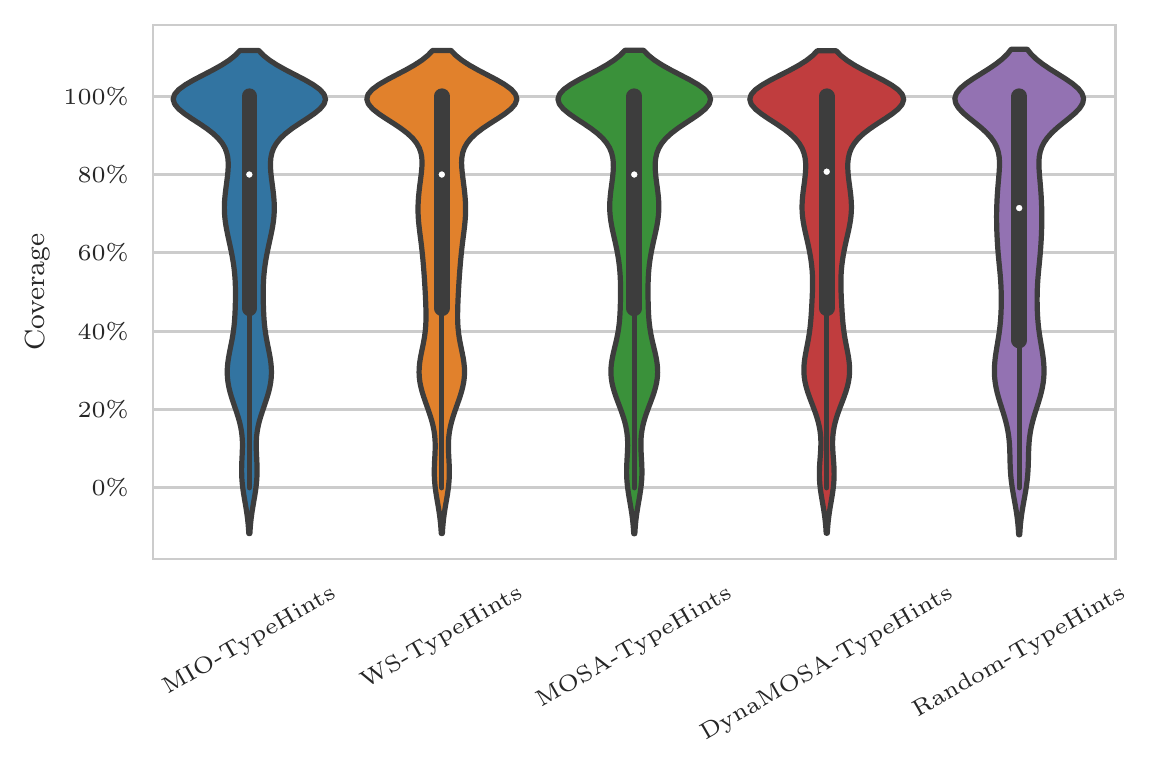}
  \caption{\label{fig:coverage-distribution}%
    Coverage distribution per algorithm with type information.
    The median value is indicated by a white dot
    within the inner quartile markers.
  }
\end{figure}

The violin plots in \cref{fig:coverage-distribution}
show the coverage distributions
for each algorithm.
We use all algorithms here
in a configuration that incorporates type information.
We note coverage values over the full range of \SIrange{0}{100}{\percent}.
Notably,
all violins show a higher coverage density
above \SI{20}{\percent}, and very few modules result in lower coverage;
this is caused by what we call import coverage.
Import coverage is achieved by importing the module;
when Python imports a module it executes all statements at module level,
such as imports,
or module-level variables.
It also executes all function definitions~(the \texttt{def} statement
but not the function's body or any closures)
as well as class definitions
and their respective method definitions.
Due to the structure of the Python bytecode
these definitions are also (branchless) coverage targets
that get executed anyway.
Thus,
they count towards coverage of a module.
As a consequence coverage cannot drop below import coverage.
%

The distributions for the different configurations look very similar,
indicating a very similar performance characteristics of the algorithms;
the notable exception is the Random algorithm
with a lower performance compared to the evolutionary algorithms.

Although the violin plot reports the median values~(indicated
by a white dot),
we additionally report the median and mean coverage values
for each configuration
in \cref{tab:coverage-comparison}.
The table shows the almost equal performance
of the evolutionary algorithms DynaMOSA, MIO, and MOSA.
Our random algorithm achieves the lowest coverage values
in this experiment.
\begin{table}[th]
  \tabcaption{\label{tab:coverage-comparison}%
    Comparison of the different algorithms,
    with and without type information.
    The table shows the median coverage value,
    as well as the mean coverage value with standard deviation.
  }
  \centering
  \begin{tabular}{@{} l S[table-number-alignment=right,round-mode=places,round-precision=1] r S[table-number-alignment=right,round-mode=places,round-precision=1] r @{}} \toprule
Algorithm & \multicolumn{2}{c}{With Type Hints} & \multicolumn{2}{c}{Without Type Hints} \\
 & {median (\%)} & {mean (\%)} & {median (\%)} & {mean (\%)} \\ \midrule
DynaMOSA & 80.7207498383969 & \(\num[round-mode=places,round-precision=1]{71.59005093832216} \pm  \num[round-mode=places,round-precision=1]{30.501601017692202}\) & 76.47058823529412 & \(\num[round-mode=places,round-precision=1]{69.40724385009968} \pm  \num[round-mode=places,round-precision=1]{31.024142875254217}\) \\
MIO & 80.0 & \(\num[round-mode=places,round-precision=1]{71.3231377723311} \pm  \num[round-mode=places,round-precision=1]{30.72702253939233}\) & 75.0 & \(\num[round-mode=places,round-precision=1]{68.43210778861251} \pm  \num[round-mode=places,round-precision=1]{31.34696541514621}\) \\
MOSA & 80.0 & \(\num[round-mode=places,round-precision=1]{71.28297523262947} \pm  \num[round-mode=places,round-precision=1]{30.819930868767937}\) & 76.74418604651163 & \(\num[round-mode=places,round-precision=1]{68.73565466000007} \pm  \num[round-mode=places,round-precision=1]{31.651875735181083}\) \\
Random & 71.42857142857143 & \(\num[round-mode=places,round-precision=1]{66.93970907666117} \pm  \num[round-mode=places,round-precision=1]{31.508764298473306}\) & 66.66666666666666 & \(\num[round-mode=places,round-precision=1]{62.622772761835066} \pm  \num[round-mode=places,round-precision=1]{32.90571525287129}\) \\
WS & 80.0 & \(\num[round-mode=places,round-precision=1]{70.63297669822799} \pm  \num[round-mode=places,round-precision=1]{30.714453301022658}\) & 71.42857142857143 & \(\num[round-mode=places,round-precision=1]{67.50810878730078} \pm  \num[round-mode=places,round-precision=1]{31.487199600297462}\) \\
\bottomrule \end{tabular}
\end{table}

Since those coverage values are so close together,
we computed \effectsize statistics for each pair of
DynaMOSA and one of the other algorithms
on the coverage of all modules.
All effects are negligible but in favour of DynaMOSA~(DynaMOSA and MIO:
\(\effectsize = \num{\MIOTypeHintsEffect}\);
DynaMOSA and MOSA: \(\effectsize = \num{\MOSATypeHintsEffect}\);
DynaMOSA and Random: \(\effectsize = \num{\RandomTypeHintsEffect}\);
DynaMOSA and WS: \(\effectsize = \num{\WSTypeHintsEffect}\)).
The effects are not significant except for DynaMOSA and Random
with \(p = \num{\RandomTypeHintsPValue}\).
We also compared the effects on coverage on a module level.
The following numbers report the count of modules
where an algorithm performed significantly better or worse
than DynaMOSA.
MIO performed better than DynaMOSA on
\num[round-precision=1]{\CoverageImprovementsMIOWorse} modules
but worse on \num[round-precision=2]{\CoverageImprovementsMIOBetter}.
MOSA performed better than DynaMOSA on
\num[round-precision=1]{\CoverageImprovementsMOSAWorse} module
but worse on \num[round-precision=1]{\CoverageImprovementsMOSABetter}.
Random performed better than DynaMOSA on
\num[round-precision=1]{\CoverageImprovementsRandomWorse} modules
but worse on \num[round-precision=2]{\CoverageImprovementsRandomBetter}.
Whole Suite performed better than DynaMOSA on
\num[round-precision=1]{\CoverageImprovementsWSWorse} module
but worse on \num[round-precision=2]{\CoverageImprovementsWSBetter}.
We see that although modules exist
where other algorithms outperform DynaMOSA significantly,
overall DynaMOSA performs better than the other algorithms.
%

\begin{figure}[th]
  \centering
  \includegraphics[]{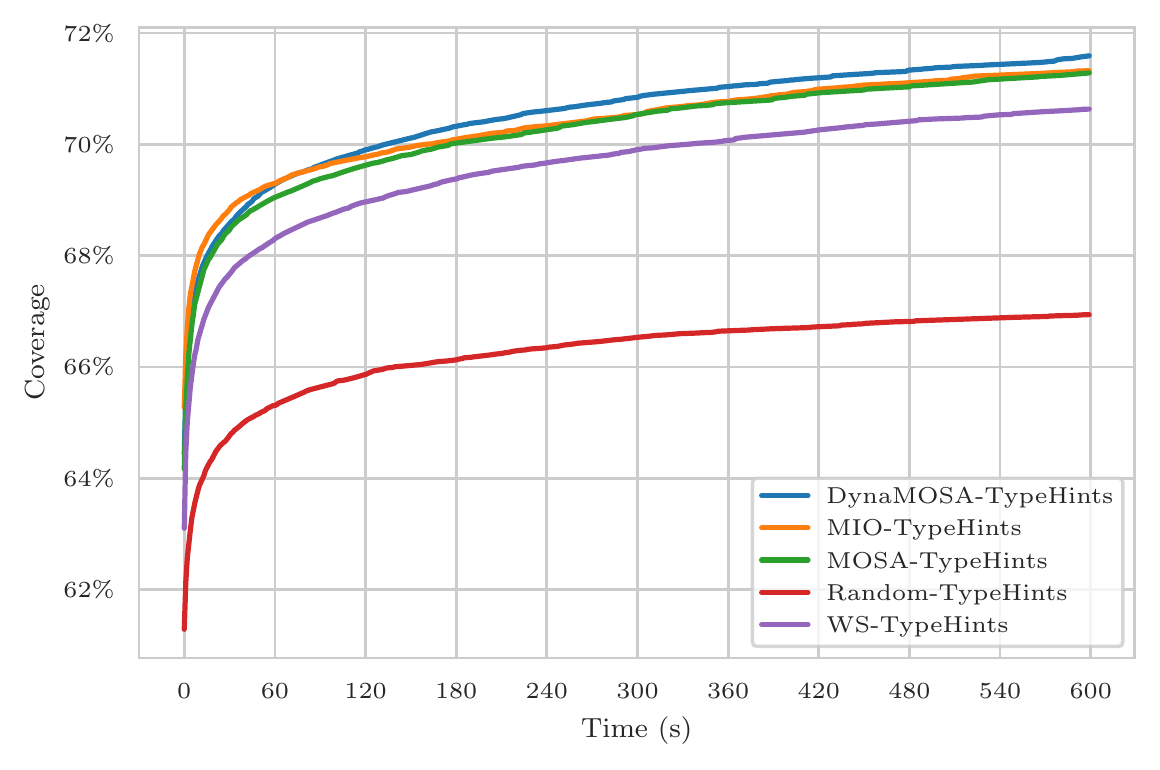}
  \caption{\label{fig:coverage-over-time-with-types}%
    Development of the coverage over time with available type information.
  }
\end{figure}

We now show the development of the coverage
over the full generation time of \SI{600}{\second}.
The line plot in \cref{fig:coverage-over-time-with-types}
reports the mean coverage values per configuration
measured in one-second intervals.
We see
that during the first minute,
MIO yields the highest coverage values
before DynaMOSA is able to overtake MIO,
while MOSA can come close.
However,
the performance of MIO decreases over the rest of the exploration phase.
From the plot we can see that MOSA comes close to MIO
at around \SI{300}{\second}.
At this point,
MIO switches over to its exploitation phase,
which again seems to be beneficial compared to MOSA.
Over the full generation time,
Whole Suite yields smaller coverage values
than the previous three,
as does Random.

\begin{figure}[th]
  \centering
  \includegraphics[]{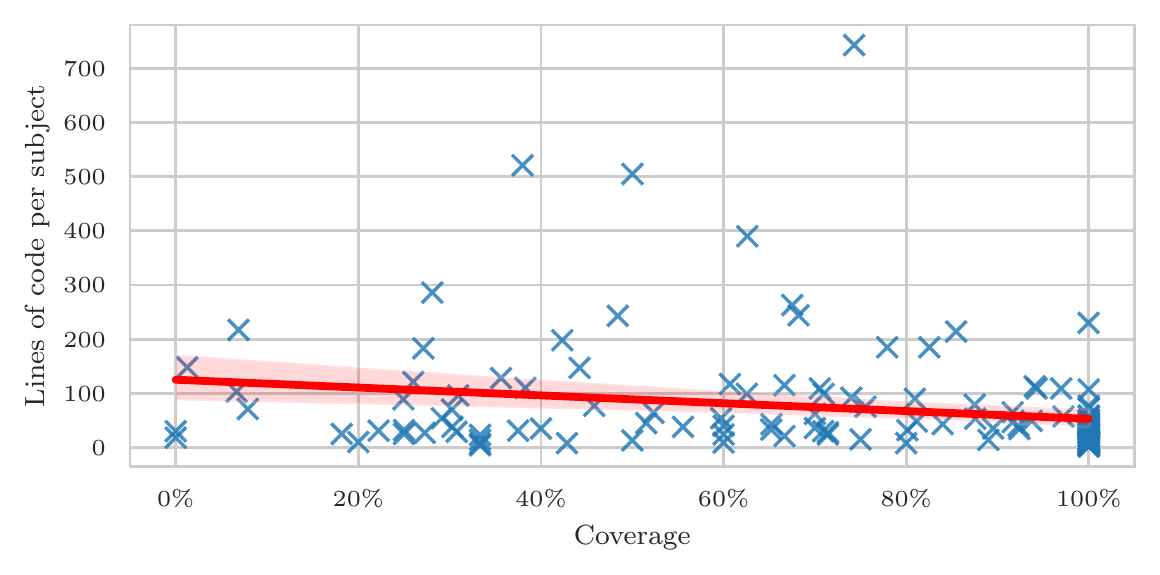}
  \caption{\label{fig:coverage-loc-correlation-dynamosa-with-types}%
    Mean coverage for DynaMOSA with type information
    correlated to the lines of code per module;
    the red line shows the linear regression fitted to the data~(%
    Pearson~\(r = \num{\CoverageLocCorrelationDynaMOSAWithTypesPearsonR}\),
    \(p = \num{\CoverageLocCorrelationDynaMOSAWithTypesPValue}\)).
  }
\end{figure}

\begin{figure}[th]
  \centering
  \includegraphics[]{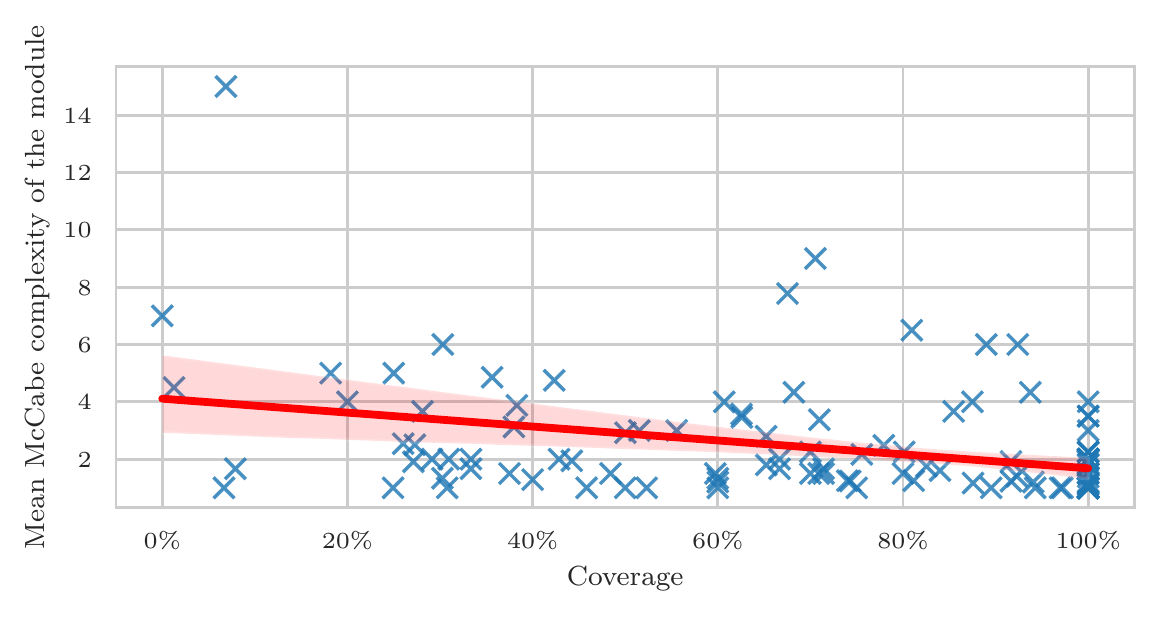}
  \caption{\label{fig:coverage-mccabe-correlation-dynamosa-with-types}%
    Mean coverage for DynaMOSA with type information
    correlated to the mean McCabe cyclomatic complexity of a module's functions;
    the red line shows the linear regression fitted to the data~(%
    Pearson~\(r = \num{\CoverageMcCabeCorrelationDynaMOSAWithTypesPearsonR}\),
    \(p = \num{\CoverageMcCabeCorrelationDynaMOSAWithTypesPValue}\)).
  }
\end{figure}

We hypothesize
that the achieved coverage is influenced by properties
of the module under test.
Our first hypothesis is
that there exists some correlation
between the achieved coverage on a module
and the number of lines of code in that module.
In order to study this hypothesis,
we use our best-performing algorithm,
DynaMOSA,
and compare the mean coverage values per module
with the lines of code in the module.
The scatter plot in \cref{fig:coverage-loc-correlation-dynamosa-with-types}
shows the result;
we fitted a linear regression line in red colour into this plot.
The data shows a weak negative correlation~(Pearson~\(r
= \num{\CoverageLocCorrelationDynaMOSAWithTypesPearsonR}\) with a \pvalue
of \num{\CoverageLocCorrelationDynaMOSAWithTypesPValue},
which indicates that there is at least some support for this hypothesis:
it is slightly easier to achieve higher coverage values
on modules with fewer lines of code.
However, lines of code is often not considered a good metric
to estimate code complexity.
Therefore,
we similarly study the correlation of mean coverage values per module
with the mean McCabe cyclomatic complexity~\citep{McC76} of that module.
The scatter plot in \cref{fig:coverage-mccabe-correlation-dynamosa-with-types}
shows the results;
again, we fitted a linear regression line in red colour into the plot.
The data shows a medium negative correlation~(Pearson~\(r
= \num{\CoverageMcCabeCorrelationDynaMOSAWithTypesPearsonR}\) with a \pvalue of
\num{\CoverageMcCabeCorrelationDynaMOSAWithTypesPValue}),
supporting this hypothesis:
modules with higher mean McCabe cyclomatic complexity
tend to be more complicated to cover.
However, since this correlation still is not strong,
other properties of a module
appear to influence the achieved coverage.
Possible properties might be the quality of available type information
or the ability of the test generator
to instantiate objects of requested types properly.
Also finding appropriate input values for function parameters
might influence the achievable coverage.
We study the influence of type information in RQ2, and
leave exploring further factors as future work.

\summary{RQ\ref{rq:quality}}{%
  Our experiments show that test-generation algorithms in Python
  yield reasonable coverage values between
  \SIrange{66.9}{71.6}{\percent} in the mean.
  Furthermore,
  we show that DynaMOSA performed best in this experiment,
  followed by MIO, MOSA, and Whole Suite;
  Random achieves the least coverage values.
}

\paragraph{Discussion:}
Overall, the results we achieved from our experiments indicate
that automated test generation for Python is feasible.
They also show
that there exists a ranking between the performance of the studied algorithms,
which is in line with previous research
on test generation in Java~\citep{CGA+18}.
%

The results,
however,
do not show large differences between the algorithms;
only DynaMOSA compared to Random yielded a significant,
although negligible,
effect.
A reason,
why the algorithms have very similar performance
is indicated by the subjects that we use for our study.
We noticed subjects
that are trivial to cover for all algorithms.
\Cref{lst:example-flutes-math} shows such an example
taken from the module \texttt{flutes.math}
from the \texttt{flutes} project\footnote{%
  \url{https://www.pypi.org/project/flutes}, accessed 2022–07–14.%
}.
\begin{lstlisting}[%
  float=th,%
  caption={A function that can be trivially covered, taken from
  \texttt{flutes.math}.},%
  label={lst:example-flutes-math},%
  language=Python,%
]
def ceil_div(a: int, b: int) -> int:
    r"""Integer division that rounds up."""
    return (a - 1) // b + 1
\end{lstlisting}
Almost every pair of two integers is a valid input
for this function~(only \texttt{b = 0} will cause a \texttt{ZeroDivisionError}).
Since this function is the only function in that particular module,
achieving full coverage on this module is also trivially possible
for all test-generation algorithms,
especially since they know from the type hints
that they shall provide integer values as parameters.
%

Another category of modules that is hard to cover,
independent of the used algorithm,
is due to technical limitations of \pynguin~(see \cref{sec:approach-limitations}).
Consider the minimised code from the \texttt{flutes}\footnote{%
  \url{https://www.pypi.org/project/flutes}, accessed 2022–07–14.%
} project in \cref{lst:example-flutes-timing}.
\begin{lstlisting}[%
  float=th,%
  caption={A function that is actually a context manager and a generator and
  thus cannot be covered by \pynguin due to technical limitations, taken from
  \texttt{flutes.timing}.},%
  label={lst:example-flutes-timing},%
  language=Python,%
]
import contextlib
import time

@contextlib.contextmanager
def work_in_progress(desc: str = "Work in progress"):
    print(desc + "... ", end='', flush=True)
    begin_time = time.time()
    yield
    time_consumed = time.time() - begin_time
    print(f"done. ({time_consumed:.2f}s)")
\end{lstlisting}
This function is actually both a context manager~(due to its decorator)
and a generator~(indicated by the \texttt{yield} statement).
\Pynguin currently supports neither;
the context manager would require a special syntax construct
to be called,
such as a \texttt{with} block.
Calling a function with a \texttt{yield} statement in it
does not actually execute its code.
It generates an iterator object pointing to that function.
The code will only be executed when iterating over the generator in a loop
or by explicitly calling \texttt{next} on the object.
A test case
\pynguin can come up with is similar to the one
shown in \cref{lst:example-flutes-timing-test}.
\begin{lstlisting}[%
  float=th,%
  caption={A test case from \pynguin for the function in
  \cref{lst:example-flutes-timing}.},%
  label={lst:example-flutes-timing-test},%
  language=Python,%
]
import flutes.timing as module_0

def test_case_0:
    generator_context_manager_0 = module_0.work_in_progress()
    assert generator_context_manager_0.args == ()
    assert generator_context_manager_0.kwds == {}
\end{lstlisting}
This test case would only result in a coverage of \SI{50}{\percent},
which is only import coverage
resulting from executing the \texttt{import} and \texttt{def} statements
during module loading.
As stated above,
the body of the function will not even be executed.
%

\Cref{fig:coverage-loc-correlation-dynamosa-with-types}
and \cref{fig:coverage-mccabe-correlation-dynamosa-with-types}
indicate that our subject modules are not very complex.
Previous research has shown
that algorithms like DynaMOSA or MIO
are more beneficial for a large number of goals,
that is,
a large number of branches~\citep{PKT15,Arc18}.
For modules with only few branches
they cannot show their full potential
which definitely influences our results.
Having smaller modules is a property of our evaluation set.
On average,
our modules consist of \num{79.21472} lines of code;
\citet{MLG21} report an average module size of \num{119.6198} lines of code.
Future work shall repeat our evaluation
using more complex subject systems
in order to evaluate
whether the assumed improvements
can be achieved there.
%


\subsection{RQ\ref{rq:typeinfluence}: Influence of Type Information}%
\label{sec:evaluation-rq2}
%

We hypothesized in the previous section
that type information might have an impact
on the achieved coverage.
We compare the configuration with type information
and the configuration without type information
for each algorithm.
Our comparison is done on the module level.
%

\begin{figure}[th]
  \centering
  \includegraphics[]{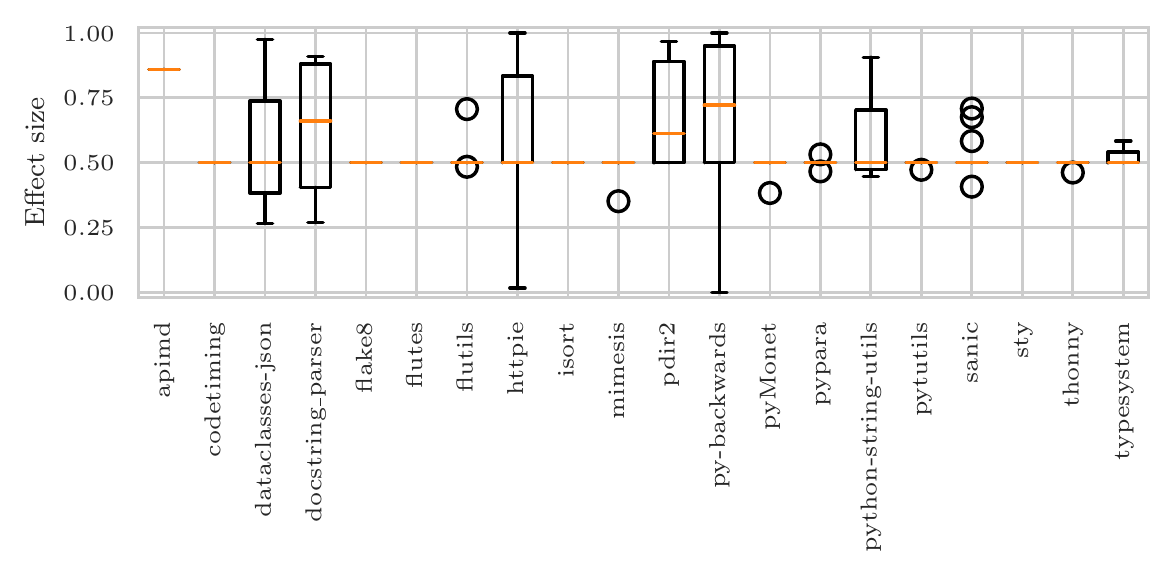}
  \caption{\label{fig:effects-dynamosa-with-without-types}%
    Effect-size distributions per project for DynaMOSA.
    \(\effectsize > \num{0.5}\) indicates that DynaMOSA-TypeHints
    performed better than DynaMOSA-NoTypeHints;
    vice versa for \(\effectsize < \num{0.5}\).
  }
\end{figure}

We plot the effect-size distributions per project for DynaMOSA,
our best-performing algorithm from RQ\ref{rq:quality},
in \cref{fig:effects-dynamosa-with-without-types}.
We aggregate the data for a better overview here;
we will also show the data on a module level afterwards.
Each data point that is used for the plot
is the effect size on one module of that project.
The results show that the median effect is always
greater or equal than \num{0.5}.
This entails that available type information
is beneficial in the mean
for four out of our twenty projects.

We do not only report the data on a project level
as we did using \cref{fig:effects-dynamosa-with-without-types}
but also per module.
We provide a table that reports
the effect size per module
in \cref{tab:improvement of types} in the appendix.
A value larger than \num{0.5} for a module
indicates that the configuration with type information
yielded higher coverage results
than the configuration without type information.
A value smaller than \num{0.5} indicates the opposite.
We use a bold font to mark a table entry
where the effect size is significant
with respect to a \(\pvalue < 0.05\).
%

\summary{RQ\ref{rq:typeinfluence}}{%
  Our experiment shows
  that type information can be beneficial
  for the effectiveness of the test-generation algorithms.
  The effect,
  however,
  largely depends on the module under test.
}

\paragraph{Discussion:}
Our results show that type information can be beneficial
to achieve higher coverage,
although this largely depends on the module under test.
From \cref{tab:improvement of types} we also note
that there are cases
where the presence of type information
has a negative impact
compared to its absence.
%

A reason for this is
that the type information is annotated manually
by the developers of the projects.
Although
powerful static type checking tools,
such as \toolname{MyPy} are available,
they are often not used.
\citet{RMM+20} conducted a study using more than \num{2600} repositories
consisting of more than \num{173000} files
that have partial type annotations available.
They report that often type information is only partially available.
As a consequence,
large parts of the code still do not have any type information at all.
They furthermore report
that the type annotations are often wrong, too:
they were able to correctly type check
only about \SI{11.8745}{\percent} of the repositories
using the \toolname{MyPy} tool.
Such wrong or incomplete type information
can mislead \pynguin as to which parameter types to use.
For wrong type information
there is still a small probability for \pynguin
to disregard the type hint anyway and use a random type.
The consequence for missing type information is
that \pynguin needs to use a random type, anyway.
%

We also noticed cases
where parts of the code can only be executed
if one uses a type for an argument
that is different from the annotated type.
An example can be error-handling code
that explicitly raises an error
in case the argument type is wrong.
We depict a simple example of such a case in \cref{lst:example-precondition}.
\begin{lstlisting}[
  float=th,%
  caption={A function having a precondition.  The then-branch can only be
  executed if one uses a type for the argument that is different from the
  annotated type.},%
  label={lst:example-precondition},%
  language=Python,%
]
def example(a: int) -> int:
    if isinstance(a, str):
        raise TypeError("Expected an int but not a string")
    return 2 * a
\end{lstlisting}
In this example,
we see that the then-branch of the \texttt{isinstance} check
can only be executed if the argument to the function
is of type \texttt{str}—in contradiction to the annotated \texttt{int}.
Although it is possible
that \pynguin disregards the type hint
there is only a small probability for this.
Besides,
\pynguin would then pick a random type from all available types
in the test cluster,
where again the probability that it picks a \texttt{str} is small,
depending on the size of the test cluster.
%

Yet another reason for the results
lies in the limitations
\pynguin currently has with respect to generating
objects of specific characteristics.
\Pynguin is,
for example,
not able to provide generators or iterators as arguments;
neither does is provide higher-order functions.
Adding these features is open as future work.
Higher-order functions are required as an argument
to \num[round-precision=2]{49} functions throughout the
\num[round-precision=3]{\numUniqueClasses} modules.
Being able to generate higher-order functions
in the context of a dynamically-typed programming language
has also been shown to be beneficial~\citep{SPK+18};
we leave this for future work.
%


\subsection{RQ\ref{rq:assertions}: Assertion Effectiveness to Reveal Faults}%
\label{sec:evaluation-rq3}

\begin{figure}[th]
  \centering
  \includegraphics[]{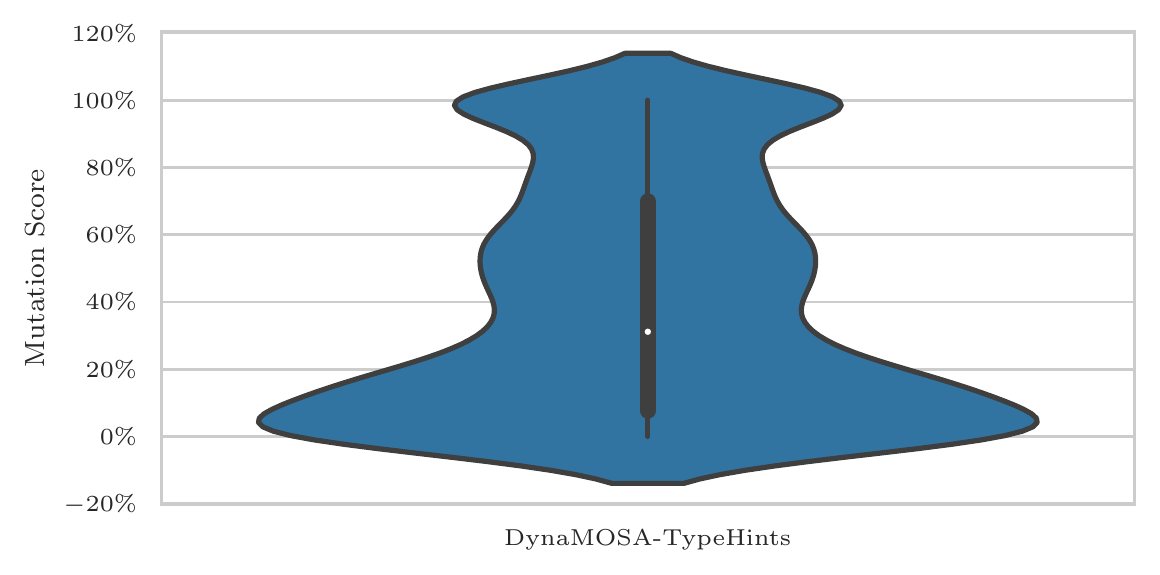}
  \caption{\label{fig:mutation-score-distribution}%
    A violinplot showing the distribution of the mutation scores
    for the used DynaMOSA configuration
    in our experiment for RQ\ref{rq:assertions}.
  }
\end{figure}

For our last research question
we study the assertion effectiveness.
To measure the effectiveness of the generated assertions
we use the mutation-score metric.
We limit our presentation of results for this questions
to a smaller subset of our modules:
\num[round-precision=3]{\numUniqueClassesWithAssertions} modules
that we ran with DynaMOSA-TypeHints,
our best-performing configuration with respect to achieved coverage.
The remaining modules from the original
\num[round-precision=3]{\numUniqueClasses}
caused various issues
and failed to be yield results for all \num[round-precision=2]{\numIterations}
reruns.
Please note that for answering RQ\ref{rq:quality} and RQ\ref{rq:typeinfluence},
we configured \pynguin in a way
that it does not create assertions
but only reaches for coverage.
We did this both to save computation time
as well as to prevent \pynguin from failing on modules
due to issues related to the assertion generation,
such as flaky tests.
The issues that cause a drop in the number of modules
for answering this research question
were caused by our used mutation approach,
flakiness,
or incorrect programs and non-terminating loops.
\Cref{fig:mutation-score-distribution} reports the distribution
of the mutation scores for this configuration in a violinplot.

\begin{figure}[th]
  \centering
  \includegraphics[]{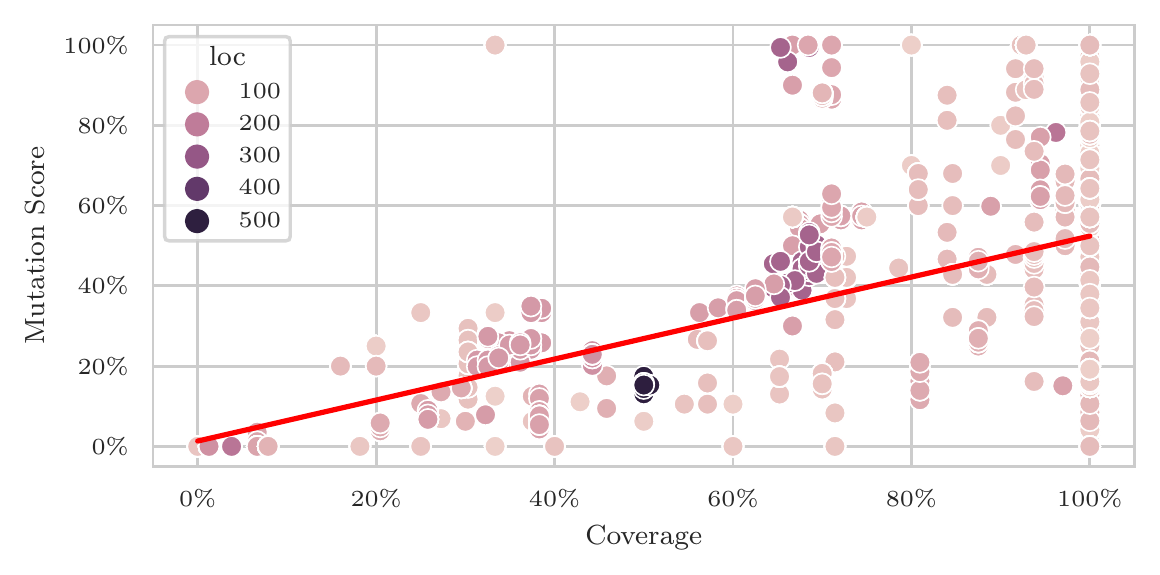}
  \caption{\label{fig:coverage-mutation-score-correlation}%
    Correlation between coverage and mutation score for DynaMOSA-TypeHints~(%
    Pearson~\(r = \num{\CoverageMutationScoreCorrelationDynaMOSAWithTypesPearsonR}\),
    \(p = \num{\CoverageMutationScoreCorrelationDynaMOSAWithTypesPValue}\)).
  }
\end{figure}

Additionally,
we hypothesize that higher mutation scores
are correlated with higher coverage, as it
seems unlikely
that for modules with small coverage values
we can achieve high mutation scores.
The rationale behind this hypothesis is
that only mutated statements of a program
that are covered by the execution
can be detected.
For this,
we plot the achieved coverage values and mutation scores
into a scatter plots;
\cref{fig:coverage-mutation-score-correlation}
shows these results.
Besides,
we add the number of lines of code per module
as an additional dimension to the plots,
visualised by the hue;
a lighter colour stands for few lines of code,
while a dark colour symbolises a module
with a large number of code lines.

The plot shows a strong correlation between coverage and mutation score,
with Pearson's correlation coefficient being computed to be
\(r = \num{\CoverageMutationScoreCorrelationDynaMOSAWithTypesPearsonR}\)
with a \pvalue of \num{\CoverageMutationScoreCorrelationDynaMOSAWithTypesPValue}
Interestingly,
there exist a few outliers
where mutation score is \SI{100}{\percent}
although the coverage is only at a mediocre level of around \SI{30}{\percent}.
We investigated these cases
and found out that mutants were only introduced in those parts of the module
that were also covered by the test case.
We conjecture this is caused by the mutation code
which our implementation uses from the MutPy library,
in combination with limitations of \pynguin.
An example is the simplified snippet from \texttt{pyMonet}'s\footnote{%
  \url{https://pypi.org/project/pyMonet}, accessed 2022–07–15.%
} module \texttt{pymonet.task},
which we show in \cref{lst:example-pymonet}.
\begin{lstlisting}[%
  float=th,%
  caption={Simplified snipped from \texttt{pymonet.task} showing different
  results in the original version than in a mutant},
  label={lst:example-pymonet},
  language=Python,%
]
class Task:
  def __init__(self, fork):
    self.fork = fork

  @classmethod
  def of(cls, value):
    return Task(lambda _, resolve: resolve(value))

  @classmethod
  def reject(cls, value):
    return Task(lambda reject, _: reject(value))

  def map(self, fn):
    def result(reject, resolve):
      return self.fork(lambda arg: reject(arg), lambda arg: resolve(fn(arg)))
    return Task(result)

  def bind(self, fn):
    def result(reject, resolve):
      return self.fork(lambda arg: reject(arg), lambda arg: fn(arg).fork(reject, resolve))
    return Task(result)
\end{lstlisting}
The module consists of 15 branchless code objects,
each of which is a coverage goal\footnote{
  The module itself, the class, each method, each closure, and each lambda
  form such code objects.
}.
Of these, \pynguin is able to cover only five~(the module,
the class declaration,
the constructor declaration,
and the declarations of the \texttt{map} and \texttt{bind} methods).
The methods \texttt{of} and \texttt{reject} are not even called
as they are annotated as class methods,
which is currently not supported by \pynguin.
\Pynguin,
however,
is generating a test cases like the one shown in
\cref{lst:example-pymonet-test}.
\begin{lstlisting}[%
  float=th,%
  caption={A test cases for the snippet in \cref{lst:example-pymonet}.},%
  label={lst:example-pymonet-test},%
  language=Python,%
]
import pymonet.task as module_0

def test_case_0:
  int_0 = 317
  task_0 = module_0.Task(int_0)
  assert module_0.Task.of is not None
  assert module_0.Task.reject is not None
\end{lstlisting}
When mutating the module from \cref{lst:example-pymonet},
the only possible change is to remove a \texttt{@classmethod} decorator.
The consequence is that \texttt{of} or \texttt{reject}
become normal methods by this operation.
Executing the test case from \cref{lst:example-pymonet-test}
against such a mutant causes a difference in the result
and thus \pynguin counts the mutant as killed.
As a consequence,
both possible mutants get killed,
although only one third of the modules coverage goals are actually covered.
Furthermore,
we also note many modules
for those the tests achieve \SI{100}{\percent} coverage
but the achieved mutation scores range from \SIrange{0}{100}{\percent};
this shows that the effectiveness of our assertions
can still be improved in the future.
%

\summary{RQ\ref{rq:assertions}}{%
  Our experiment shows
  that \pynguin is able to generate assertions
  of good effectiveness—they can achieve mutation scores of up to \SI{100}{\percent}.
}

\paragraph{Discussion:}
Our experiment indicates
that \pynguin is able to provide high-quality assertions.
As we expected,
we note a strong correlation
between coverage and mutation score~(see
\cref{fig:coverage-mutation-score-correlation}).
Using mutation score as an indicator for fault-finding capabilities
is backed by the literature~(see,
for example,
the work of \citet{JJL+14} or \citet{PSY+18}).
However,
it is still open in the literature
whether there exists a strong correlation
between mutation scores and fault detection.
To answer this research question
we had to shrink the set of subject modules
from \num[round-precision=3]{\numUniqueClasses}
down to \num[round-precision=3]{\numUniqueClassesWithAssertions}.
This is caused by various influences:
we use a customised version of \toolname{MutPy}~\citep{DH14}
to generate the mutants~(see~\cref{sec:approach-assertion-generation}).
The original \toolname{MutPy} has not received an update since 2019
and is designed to work with Python versions~3.4 to~3.7\footnote{
  \url{https://github.com/mutpy/mutpy}, accessed 2022–07–12.
};
\pynguin,
however,
requires Python version~3.10 for our experiments.
Several Python internals,
such as the abstract syntax tree~(AST)
that is used by \toolname{MutPy} to mutate the original module,
received changes between Python~3.7 and~3.10.
Although we fixed the obvious issues,
our customised version of \toolname{MutPy}
is an issue for crashes:
future work shall replace this component
by our own implementation of the various mutation operators
for a more fine-grained control.
Furthermore,
mutation at the AST level can cause various problems,
for example,
incorrect programs
or non-terminating loops.
We tried to make our test execution as robust against non-termination
as we could,
still we noted
that there are open issues
the experiment infrastructure to shutdown \pynguin
hardly after a walltime of \SI{2.5}{\hour}—without reporting any results.
In some cases already the original subject module
shows non-deterministic behaviour
for example,
due to the use of random numbers.
Our assertion generation executes each generated test case twice
in random order
to remove trivially flaky assertions.
Previous research~\citep{GLK+21},
however,
has shown that in order to achieve a \SI{95}{\percent} confidence
that a passing test case is not flaky
will require \num[round-precision=3]{170} reruns on average.
Obviously,
those reruns would consume a huge amount of time,
which \pynguin should not invest into rerunning tests
to potentially find a flaky one.
Therefore,
we accept that \pynguin generates flaky tests,
which can influence the assertion generation
and cause failures.
The results we report here,
however,
are promising.
They show that also mutation-based assertion generation
is feasible for dynamically typed programming languages;
this again opens up new research perspectives
targeting the underlying oracle problem.

\section{Related Work}\label{sec:related-work}

Closest to our work is \evosuite~\citep{FA13},
which is a unit test generation framework for Java.
Similar to our approach,
\evosuite incorporates many different test-generation strategies
to allow studies on their different behaviour and characteristics.
To the best of our knowledge, little has been done in the area
of automated test generation for dynamically typed languages.
Search-based test generation tools have been implemented before,
for example,
for the Lua programming language~\citep{WHW15}
or Ruby~\citep{MFT11}.
While these approaches utilise a genetic algorithm,
they are only evaluated on small data sets
and do not provide comparisons between different test-generation techniques.
Approaches such as \toolname{SymJS}~\citep{LAG14}
or \toolname{JSEFT}~\citep{MMP15}
target specific properties of JavaScript web applications,
such as the browser's DOM
or the event system.
Feedback-directed random testing has also been adapted to web applications
with \toolname{Artemis}~\citep{ADJ+2011}.
Recent work proposes
\toolname{LambdaTester}~\citep{SPK+18},
a test generator that specifically addresses the generation
of higher-order functions in dynamic languages.
Our approach,
in contrast,
is not limited to specific application domains.
A related approach to our random test generation
can be found in the \toolname{Hypothesis} library,
a library for property-based testing in Python~\citep{MH19,MD20}.
\toolname{Hypothesis} uses a random generator
to generate the inputs for checking whether a property holds.
It is also able to generate and export test cases
based on this mechanism.
It does,
however,
only implement a random generator
and does not focus on unit-test generation primarily.
TSTL~\citep{GP15,HGP+18} is another tool for property-based testing in Python
although it can also be used to generate unit tests.
It utilises a random test generator for this
but its main focus is not to generate unit tests automatically,
which we aim to achieve with our \pynguin approach.
Further tools are,
for example,
\toolname{Auger}\footnote{%
  \url{https://github.com/laffra/auger}, accessed 2022–07–12.%
},
\toolname{CrossHair}\footnote{
  \url{https://crosshair.readthedocs.io}, accessed 2022–02–10,%
},
or \toolname{Klara}\footnote{
  \url{https://klara-py.readthedocs.io}, accessed 2022–02–10,%
};
they all require manual effort by the developer to create test cases
in contrast to our automated generation approach.
Additionally,
dynamically typed languages have also gained attention
from research on test amplification.
\citet{SAD21} introduce \toolname{PyAmplifier}
a proof-of-concept tool for test amplification in Python.
Recently,
\toolname{Small-Amp}~\citep{ARD+22}
demonstrated the use of dynamic type profiling
as a substitute for type declarations
in the Pharo Smalltalk language.
In recent work,
\pynguin has also been evaluated on a
scientific software stack from material sciences~\citep{TMG22}.
The authors also compared the performance
of different test-generation algorithms
and the assertion generation on their subject system;
their results regarding the algorithm performance
are in line with our results,
whereas their results on assertion quality differ significantly
due to their use of an older version of \pynguin
that only provided limited assertion-generation capabilities.

\section{Conclusions}\label{sec:conclusions}

In this work,
we presented \pynguin,
an automated unit test generation framework for Python.
We extended our previous work~\citep{LKF20}
by incorporating the DynaMOSA, MIO, and MOSA algorithms, and
by evaluating the achieved coverage on a larger corpus of Python modules.
Our experiments demonstrate that \pynguin is able to emit unit tests for Python
that cover large parts of existing code bases.
In line with prior research in the domain of Java unit test generation,
our evaluation results show
that DynaMOSA performs best in terms of branch coverage,
followed by MIO, MOSA, and the Whole Suite approach.
While our experiments provide evidence
that the availability of type information influences the performance
of the test-generation algorithms,
they also show that there are several open issues
that provide opportunities for further research.
A primary challenge is
that adequate type information may not be available in the first place,
suggesting synergies with research on type inference.
However,
even if type information is available,
generating instances of the relevant types remains challenging.
There are also technical challenges
that \pynguin needs to address in order to become practically usable;
overcoming technical limitations is necessary
to allow the usage of \pynguin for a wider field of projects and scenarios.
%

While we were investigating in the technical limitations of \pynguin,
we also found a real bug
in one of our subjects:
the module \texttt{mimesis.providers.date} from the \texttt{mimesis}\footnote{%
  \url{https://pypi.org/project/mimesis}, accessed 2022–07–14.%
}
contains the following function presented in \cref{lst:mimesis-buggy}~(we
removed various lines
that are not relevant to the bug for presentation reasons).
\begin{lstlisting}[
  float=th,%
  caption={Simplified code snippet taken from \texttt{mimesis.providers.date}},%
  label={lst:mimesis-buggy},%
  language=Python,%
]
from datetime import datetime, timedelta

def bulk_create_datetimes(
    date_start: datetime, date_end: datetime, **kwargs: Any
) -> list[datetime]:
    dt_objects = []

    if date_end < date_start:
        raise ValueError()

    while date_start <= date_end:
        date_start += timedelta(**kwargs)
        dt_objects.append(date_start)

    return dt_objects
\end{lstlisting}

The \texttt{mimesis} project provides utility functions
that generate fake data,
which aims to look as realistic as possible.
The function presented in \cref{lst:mimesis-buggy}
aims to generate a list of \texttt{datetime} objects~(essentially,
a combination of date and time stamp)
between a given start and end time
using a specific interval.
It requires a start time
that is smaller than the end time,
that is,
lies before the end time.
Besides,
it takes a dictionary in its \texttt{**kwargs} parameter,
which it hands over to the \texttt{timedelta} function
of Python's \texttt{datetime} API.
%

The \texttt{timedelta} function is used to compute a delta
that can be added to the current time
in order to retrieve the next date.
Its API is very flexible:
it allows to specify an arbitrary delta,
especially also of negative and zero values.
While we were investigating in \pynguin's timeout handling,
we noticed that \pynguin generated a test case for this function
similar to the one presented in \cref{lst:mimesis-testcase}.
\begin{lstlisting}[
  float=th,%
  caption={A bug-exposing test case},%
  label={lst:mimesis-testcase},%
  language=Python,%
]
import datetime
import mimesis.providers.date as module_0

def test_case():
    int_0 = 2022
    int_1 = 1
    int_2 = 2
    date_time_0 = datetime.datetime(int_0, int_1, int_1)
    date_time_1 = datetime.datetime(int_0, int_1, int_2)
    module_0.bulk_create_datetimes(date_time_0, date_time_1)
\end{lstlisting}
Executing this test case leads to an infinite loop
because the \texttt{timedelta} function will yield a delta of zero
if no parameters given.
%

We reported this issue to the developers of \texttt{mimesis},
who stated that this \enquote{is a major bug actually}\footnote{
  \url{https://github.com/lk-geimfari/mimesis/pull/1229#issuecomment-1162974494},
  accessed 2022–07–14.%
}.
They furthermore accepted the fix proposed by the second author of this paper
and released a new version of their library.
More information on \pynguin is available from its web page:
\begin{center}
  \url{https://www.pynguin.eu}
\end{center}
%
%

%
\section*{Conflict of interest}

The authors declare that they have no conflict of interest.

\bibliographystyle{spbasic}      
\bibliography{related}   

\clearpage
\appendix

%
%

\section{Comparison of Coverage Improvements}\label{sec:appendix-coverage}

The following \cref{tab:improvement of types}
presents the Vargha-Delaney \effectsize{} statistics
for each module and algorithm on the resulting coverage values.
A value larger than \num{0.5} indicates
that the configuration with type information
yielded higher coverage results
than the configuration without type information;
vice versa for values smaller \num{0.5}.
We use a \textbf{bold} font
to indicate effect sizes
that are significant.

\begin{longtable}{@{} l r @{}}
\caption{Improvement of types. The table shows the Vargha-Delaney \effectsize{} statistic for each module and DynaMOSA. A value larger than \num{0.5} indicates that the configuration with type information yielded higher coverage results than the configuration without type information; vice versa for values smaller \num{0.5}.  In \textbf{bold} font we denote those effects that are significant at a \(\pvalue < \num{0.05}\).}\label{tab:improvement of types}\\
\toprule
Module Name & Effect Size \\ \midrule
\endfirsthead
\multicolumn{2}{@{}l}{\small\ldots\emph{continued}} \\ \toprule
Module Name & Effect Size \\ \midrule
\endhead
\bottomrule
\multicolumn{2}{r@{}}{\small\emph{continued on next page} \ldots} \\
\endfoot
\bottomrule
\endlastfoot
\texttt{apimd.compiler} & \textbf{\num{0.86}} \\
\rowcolor{gray!25}
\texttt{codetiming.\_timer} & \num{0.5} \\
\texttt{dataclasses\_json.api} & \num{0.5} \\
\rowcolor{gray!25}
\texttt{dataclasses\_json.mm} & \textbf{\num{0.975}} \\
\texttt{dataclasses\_json.undefined} & \textbf{\num{0.26666666666666666}} \\
\rowcolor{gray!25}
\texttt{docstring\_parser.google} & \textbf{\num{0.9088888888888889}} \\
\texttt{docstring\_parser.numpydoc} & \textbf{\num{0.8711111111111111}} \\
\rowcolor{gray!25}
\texttt{docstring\_parser.parser} & \num{0.45} \\
\texttt{docstring\_parser.rest} & \textbf{\num{0.2688888888888889}} \\
\rowcolor{gray!25}
\texttt{flake8.exceptions} & \num{0.5} \\
\texttt{flake8.formatting.base} & \num{0.5} \\
\rowcolor{gray!25}
\texttt{flake8.formatting.default} & \num{0.5} \\
\texttt{flake8.main.debug} & \num{0.5} \\
\rowcolor{gray!25}
\texttt{flake8.main.git} & \num{0.5} \\
\texttt{flutes.math} & \num{0.5} \\
\rowcolor{gray!25}
\texttt{flutes.timing} & \num{0.5} \\
\texttt{flutils.decorators} & \num{0.5} \\
\rowcolor{gray!25}
\texttt{flutils.namedtupleutils} & \num{0.5} \\
\texttt{flutils.packages} & \num{0.48444444444444446} \\
\rowcolor{gray!25}
\texttt{flutils.pathutils} & \textbf{\num{0.7066666666666667}} \\
\texttt{flutils.setuputils.cmd} & \num{0.5} \\
\rowcolor{gray!25}
\texttt{flutils.strutils} & \num{0.5} \\
\texttt{httpie.cli.dicts} & \num{0.5} \\
\rowcolor{gray!25}
\texttt{httpie.cli.exceptions} & \num{0.5} \\
\texttt{httpie.config} & \textbf{\num{0.8833333333333333}} \\
\rowcolor{gray!25}
\texttt{httpie.models} & \num{0.39444444444444443} \\
\texttt{httpie.output.formatters.colors} & \textbf{\num{1.0}} \\
\rowcolor{gray!25}
\texttt{httpie.output.formatters.headers} & \num{0.5} \\
\texttt{httpie.output.formatters.json} & \num{0.5} \\
\rowcolor{gray!25}
\texttt{httpie.output.processing} & \textbf{\num{1.0}} \\
\texttt{httpie.output.streams} & \textbf{\num{0.016666666666666666}} \\
\rowcolor{gray!25}
\texttt{httpie.plugins.base} & \textbf{\num{0.8666666666666667}} \\
\texttt{httpie.plugins.manager} & \num{0.55} \\
\rowcolor{gray!25}
\texttt{httpie.sessions} & \textbf{\num{0.7366666666666667}} \\
\texttt{httpie.ssl} & \num{0.5} \\
\rowcolor{gray!25}
\texttt{httpie.status} & \num{0.5} \\
\texttt{isort.comments} & \num{0.5} \\
\rowcolor{gray!25}
\texttt{isort.exceptions} & \num{0.5} \\
\texttt{isort.utils} & \num{0.5} \\
\rowcolor{gray!25}
\texttt{mimesis.builtins.base} & \num{0.5} \\
\texttt{mimesis.builtins.da} & \num{0.5} \\
\rowcolor{gray!25}
\texttt{mimesis.builtins.de} & \num{0.5} \\
\texttt{mimesis.builtins.it} & \num{0.5} \\
\rowcolor{gray!25}
\texttt{mimesis.builtins.nl} & \num{0.5} \\
\texttt{mimesis.builtins.pt\_br} & \num{0.5} \\
\rowcolor{gray!25}
\texttt{mimesis.builtins.uk} & \num{0.5} \\
\texttt{mimesis.decorators} & \num{0.5} \\
\rowcolor{gray!25}
\texttt{mimesis.exceptions} & \num{0.5} \\
\texttt{mimesis.providers.choice} & \textbf{\num{0.3516666666666667}} \\
\rowcolor{gray!25}
\texttt{mimesis.providers.clothing} & \num{0.5} \\
\texttt{mimesis.providers.code} & \num{0.5} \\
\rowcolor{gray!25}
\texttt{mimesis.providers.development} & \num{0.5} \\
\texttt{mimesis.providers.hardware} & \num{0.5} \\
\rowcolor{gray!25}
\texttt{mimesis.providers.numbers} & \num{0.5} \\
\texttt{mimesis.providers.science} & \num{0.5} \\
\rowcolor{gray!25}
\texttt{mimesis.providers.transport} & \num{0.5} \\
\texttt{mimesis.providers.units} & \num{0.5} \\
\rowcolor{gray!25}
\texttt{mimesis.shortcuts} & \num{0.5} \\
\texttt{pdir.\_internal\_utils} & \textbf{\num{0.6122222222222222}} \\
\rowcolor{gray!25}
\texttt{pdir.attr\_category} & \textbf{\num{0.8911111111111111}} \\
\texttt{pdir.color} & \num{0.5} \\
\rowcolor{gray!25}
\texttt{pdir.configuration} & \num{0.5} \\
\texttt{pdir.format} & \textbf{\num{0.9666666666666667}} \\
\rowcolor{gray!25}
\texttt{py\_backwards.conf} & \num{0.5} \\
\texttt{py\_backwards.files} & \textbf{\num{0.0}} \\
\rowcolor{gray!25}
\texttt{py\_backwards.transformers.base} & \textbf{\num{1.0}} \\
\texttt{py\_backwards.transformers.class\_without\_bases} & \textbf{\num{0.8833333333333333}} \\
\rowcolor{gray!25}
\texttt{py\_backwards.transformers.dict\_unpacking} & \textbf{\num{1.0}} \\
\texttt{py\_backwards.transformers.formatted\_values} & \textbf{\num{0.95}} \\
\rowcolor{gray!25}
\texttt{py\_backwards.transformers.functions\_annotations} & \num{0.5} \\
\texttt{py\_backwards.transformers.import\_pathlib} & \num{0.5} \\
\rowcolor{gray!25}
\texttt{py\_backwards.transformers.metaclass} & \textbf{\num{0.9166666666666666}} \\
\texttt{py\_backwards.transformers.python2\_future} & \num{0.5} \\
\rowcolor{gray!25}
\texttt{py\_backwards.transformers.return\_from\_generator} & \textbf{\num{0.9505555555555556}} \\
\texttt{py\_backwards.transformers.starred\_unpacking} & \textbf{\num{0.8133333333333334}} \\
\rowcolor{gray!25}
\texttt{py\_backwards.transformers.string\_types} & \num{0.5} \\
\texttt{py\_backwards.transformers.variables\_annotations} & \num{0.5} \\
\rowcolor{gray!25}
\texttt{py\_backwards.transformers.yield\_from} & \textbf{\num{1.0}} \\
\texttt{py\_backwards.types} & \num{0.5} \\
\rowcolor{gray!25}
\texttt{py\_backwards.utils.helpers} & \num{0.63} \\
\texttt{py\_backwards.utils.snippet} & \textbf{\num{1.0}} \\
\rowcolor{gray!25}
\texttt{pymonet.box} & \num{0.5} \\
\texttt{pymonet.immutable\_list} & \num{0.5} \\
\rowcolor{gray!25}
\texttt{pymonet.lazy} & \textbf{\num{0.38333333333333336}} \\
\texttt{pymonet.maybe} & \num{0.5} \\
\rowcolor{gray!25}
\texttt{pymonet.monad\_try} & \num{0.5} \\
\texttt{pymonet.semigroups} & \num{0.5} \\
\rowcolor{gray!25}
\texttt{pymonet.task} & \num{0.5} \\
\texttt{pymonet.validation} & \num{0.5} \\
\rowcolor{gray!25}
\texttt{pypara.accounting.generic} & \num{0.5} \\
\texttt{pypara.accounting.journaling} & \num{0.4666666666666667} \\
\rowcolor{gray!25}
\texttt{pypara.commons.errors} & \num{0.5} \\
\texttt{pypara.commons.numbers} & \num{0.5} \\
\rowcolor{gray!25}
\texttt{pypara.commons.others} & \num{0.5} \\
\texttt{pypara.commons.zeitgeist} & \num{0.5} \\
\rowcolor{gray!25}
\texttt{pypara.monetary} & \num{0.5322222222222223} \\
\texttt{pytutils.debug} & \num{0.5} \\
\rowcolor{gray!25}
\texttt{pytutils.excs} & \num{0.5} \\
\texttt{pytutils.files} & \num{0.5} \\
\rowcolor{gray!25}
\texttt{pytutils.lazy.lazy\_import} & \num{0.4727777777777778} \\
\texttt{pytutils.meth} & \num{0.5} \\
\rowcolor{gray!25}
\texttt{pytutils.path} & \num{0.5} \\
\texttt{pytutils.pretty} & \num{0.5} \\
\rowcolor{gray!25}
\texttt{pytutils.props} & \num{0.5} \\
\texttt{pytutils.python} & \num{0.5} \\
\rowcolor{gray!25}
\texttt{pytutils.pythree} & \num{0.5} \\
\texttt{pytutils.rand} & \num{0.5} \\
\rowcolor{gray!25}
\texttt{sanic.base} & \num{0.5} \\
\texttt{sanic.config} & \num{0.5} \\
\rowcolor{gray!25}
\texttt{sanic.cookies} & \num{0.4077777777777778} \\
\texttt{sanic.handlers} & \num{0.5} \\
\rowcolor{gray!25}
\texttt{sanic.headers} & \textbf{\num{0.6755555555555556}} \\
\texttt{sanic.helpers} & \num{0.5} \\
\rowcolor{gray!25}
\texttt{sanic.mixins.listeners} & \num{0.5} \\
\texttt{sanic.mixins.middleware} & \num{0.5} \\
\rowcolor{gray!25}
\texttt{sanic.mixins.routes} & \textbf{\num{0.7088888888888889}} \\
\texttt{sanic.mixins.signals} & \num{0.5} \\
\rowcolor{gray!25}
\texttt{sanic.models.futures} & \num{0.5} \\
\texttt{sanic.models.protocol\_types} & \num{0.5} \\
\rowcolor{gray!25}
\texttt{sanic.views} & \num{0.5833333333333334} \\
\texttt{string\_utils.errors} & \num{0.5} \\
\rowcolor{gray!25}
\texttt{string\_utils.manipulation} & \textbf{\num{0.9061111111111111}} \\
\texttt{string\_utils.validation} & \num{0.44722222222222224} \\
\rowcolor{gray!25}
\texttt{sty.lib} & \num{0.5} \\
\texttt{sty.register} & \num{0.5} \\
\rowcolor{gray!25}
\texttt{sty.renderfunc} & \num{0.5} \\
\texttt{thonny.languages} & \num{0.5} \\
\rowcolor{gray!25}
\texttt{thonny.plugins.pgzero\_frontend} & \num{0.5} \\
\texttt{thonny.roughparse} & \num{0.4622222222222222} \\
\rowcolor{gray!25}
\texttt{thonny.terminal} & \num{0.5} \\
\texttt{thonny.token\_utils} & \num{0.5} \\
\rowcolor{gray!25}
\texttt{typesystem.tokenize.positional\_validation} & \num{0.5} \\
\texttt{typesystem.tokenize.tokenize\_yaml} & \textbf{\num{0.5833333333333334}} \\
\rowcolor{gray!25}
\texttt{typesystem.unique} & \num{0.5} \\
\end{longtable}

\end{document}